\documentclass[prb,twocolumn,superscriptaddress,longbibliography,
 aps,floatfix]{revtex4-2}
 
\pdfoutput=1
\usepackage{graphicx}
\usepackage{dcolumn}
\usepackage{bm}
\usepackage{amssymb}
\usepackage{amsmath}
\usepackage{float}
\usepackage{enumitem}
\usepackage[colorlinks=true, citecolor=blue, urlcolor=blue]{hyperref}
\newcommand{\be}{\begin{equation}}
\newcommand{\ee}{\end{equation}}

\begin{document}

\preprint{APS/123-QED}

\title{Exact solution for the filling-induced thermalization transition in a 1D fracton system}

\author{Calvin Pozderac}
\affiliation{Department of Physics, Ohio State University, Columbus, Ohio 43210, USA}
\author{Steven Speck}
\affiliation{Department of Physics, Ohio State University, Columbus, Ohio 43210, USA}
\author{Xiaozhou Feng}
\affiliation{Department of Physics, Ohio State University, Columbus, Ohio 43210, USA}
\author{David A. Huse}
\affiliation{Department of Physics, Princeton University, Princeton, NJ 08544, USA}
\author{Brian Skinner}
\affiliation{Department of Physics, Ohio State University, Columbus, Ohio 43210, USA}

\date{\today}% It is always \today, today,

\begin{abstract}

We study a random circuit model of constrained fracton dynamics, in which particles on a one-dimensional lattice undergo random local motion subject to both charge and dipole moment conservation. The configuration %Hilbert 
space of this system exhibits a continuous phase transition between a weakly fragmented (``thermalizing'') phase and a strongly fragmented (``nonthermalizing'') phase as a function of the number density of particles. Here, by mapping to two different problems in combinatorics, we identify an exact solution for the critical density $n_c$. Specifically, when evolution proceeds by operators that act on $\ell$ contiguous sites, the critical density is given by $n_c = 1/(\ell -2)$. We identify the critical scaling near the transition, and we show that there is a universal value of the correlation length exponent $\nu = 2$. We confirm our theoretical results with numeric simulations.  In the thermalizing phase the dynamical exponent is subdiffusive: $z=4$, while at the critical point it increases to $z_c \gtrsim 6$.
%which suggest that the dynamical exponent $z$ becomes large near the transition.

%We study a 1D fracton lattice model under the influence of random dipole-conserving local gates of size $\ell$. We calculate an exact critical filling, $n_c = 1/(\ell-2)$, that separates the strongly and weakly fragmented regimes. Below $n_c$, the Hilbert space is strongly shattered into measure zero disconnected subsectors in which the fractons become localized.  Above $n_c$, the system is only weakly fragmented and thus the system can subdiffusively thermalize since nearly all of the possible configurations with a given particle number, $N$, and dipole moment, $P$, are dynamically connected. The dynamical exponent in the thermalized regime is found to be the expected $z=4$ and is larger exactly at the critical filling. The value of $n_c$ is determined exactly through analogy with integer lattice paths and tournament scoring sequences and corroborated by numerical results. Approaching this critical filling, we also calculate a correlation length of locally thermalized regions with exponent $\nu=2$.

\end{abstract}

\maketitle

\section{Introduction}
\label{sec:introduction}

An isolated system with many degrees of freedom is {\it thermalizing} if it is able to dynamically act as a bath for all of its small subsystems and thus bring them all to thermal equilibrium with each other.
%The fundamental postulate of thermodynamics is that an isolated system occupies all microstates that are consistent with its conserved quantities with equal probability. Where this postulate is obeyed, an isolated system can be said to be \emph{thermalized}.
The eigenstate thermalization hypothesis (ETH) extends these considerations %this postulate 
to specific quantum states, by asserting that when a large thermalizing system is in an energy eigenstate, the reduced density operator of each of its small subsystems is the same as in the corresponding standard thermal ensemble %state itself is thermalized if any macroscopic subsystem is well described as a thermal ensemble with the same energy density 
(see, e.g., Refs.~\cite{Nandkishore2015, deutsch_eigenstate_2018} for reviews of ETH).
The last few decades have seen intense interest in systems and states that fail to thermalize or to obey the ETH, and which therefore cannot be described by conventional equilibrium thermodynamics even at arbitrarily long times.  Some prominent examples include many-body localized states \cite{Anderson58, Basko-Altshuler2006, Pal-Huse2010, Oganesyan-Huse2007, Prelovsek, Imbrie2016, mblrmp, Pal-Huse2010} and quantum scar states \cite{Heller_bound_1984, ShiraishiMori, AKLT1, Turner-Papic2018a,Turner-Papic2018b,Ho-Lukin2019,Khemani-Chandran2019,Lin-Motrunich2019,Pancotti-Banuls2019,SchecterXY,Iadecola,LesikTower, SanjayTower,Moudgalya-Regnault2019,Moudgalya-Bernevig2020,Mark-Motrunich2020a}. 

The recently-identified \emph{fracton} systems \cite{Chamon2005, BRAVYI2011839, Haah2011, Vijay_Haah_Fu_2015, Vijay_Haah_Fu_2016, Pretko_2017, gromov_classification_2019, Doshi2021, nandkishore_review_2019, pretko_review_2020} provide yet another pathway by which a system can fail to thermalize. In fracton systems, thermalization can be avoided because of kinetic constraints on the system's dynamics, which prevent the system from exploring the full set of states consistent with the conserved quantities. 
A now-paradigmatic example of fracton dynamics is that of a one-dimensional system of charges for which both the charge and dipole moment are conserved. The dipole moment conservation ensures that a single, isolated charge cannot move freely through the system, unless an opposite-facing dipole is simultaneously created from the vacuum \cite{Pretko_2017, pretko_gauge_principle_2018}. Recent work has shown that when such a system evolves under local dynamics, the configuration space associated with a given symmetry sector can become ``fragmented'' \cite{khemani_shattering_2020, Rakovszky_Sala_Verresen_Knap_Pollmann2020,  Sala2020, moudgalya_thermalization_2021, Moudgalya_Hilbert_2022}. That is, the set of all microstates that are consistent with a given value of charge and dipole moment may separate into many dynamically disconnected sectors which are mutually inaccessible by the dynamics. Here we refer to these dynamically disconnected sectors as ``Krylov sectors.''

Fragmentation of the symmetry sector, where it occurs, may happen in either a \emph{weak} or a \emph{strong} way \cite{Sala2020, khemani_shattering_2020}. Under weak fragmentation, there is a dominant Krylov sector that contains the vast majority of states in the symmetry sector, such that in the limit of infinite system size the probability that a randomly-chosen state is contained within the largest Krylov sector approaches unity. When there is strong fragmentation, on the other hand, even the largest Krylov sector contains a vanishingly small fraction of the symmetry sector.  If we assume that the dynamics is ergodic within each Krylov sector, then in the latter case no initial state is able to thermalize, while in the case of weak fragmentation a randomly-chosen initial state will, with a probability that approaches unity in the thermodynamic limit, thermalize. Thus, as a shorthand, throughout this paper we refer to the transition between strong and weak fragmentation as the ``thermalization transition''.

Initial work on the thermalization transition in fracton systems focused on the effect of varying the spatial range $\ell$ of the operators governing dynamical evolution, or on varying the size $q$ of the local Hilbert space at each site \cite{Pai-Nandkishore2019, PhysRevX.10.011047,PhysRevE.103.022142, PhysRevLett.125.245303, Moudgalya_Prem_Huse_Chan2021}. When $\ell$ or $q$ is large enough, the system thermalizes under either random dynamics or certain types of Hamiltonian dynamics, while small $\ell$ and $q$ prevents thermalization. In a recent paper, however, Morningstar et.~al.~\cite{PhysRevB.101.214205} showed that the thermalization transition may also be effected by changing the \emph{filling} of the system for fixed $\ell$ and $q$. Here, as an example, we focus on the case of a one-dimensional lattice of sites for which the charge at each site %(which we treat as a bosonic occupation variable) 
can be any non-negative integer. If the average filling $n$ of the lattice satisfies $n \ll 1$, then a typical %basis
state consists of rare charges that are well separated from each other. Since isolated charges are unable to move while satisfying the dipole moment constraint (and since negative values of the charge are forbidden), this system is unable to evolve under the action of local operators, and it fails to thermalize. On the other hand, when $n \gg 1$ local operators can easily rearrange charges locally while satisfying the dipole moment constraint, and the system thermalizes. Thus, varying the filling $n$ allows one to study the thermalization transition in terms of a continuous variable (unlike $\ell$ and $q$, which are discrete), and thus to identify the critical exponents and critical scaling associated with the transition.

In this paper, we focus on the model introduced in the previous paragraph (which we define more precisely below), which differs slightly from that of Ref.~\cite{PhysRevB.101.214205}, and we study the filling-induced thermalization transition. In addition to numeric simulations, we provide exact solutions for the size of the symmetry sector and also the size of what appears to be the largest Krylov sector in the large-system limit. These solutions, which we obtain by a mapping to two separate problems in combinatorics, provide us with exact solutions for the critical filling $n_c$ as a function of gate size $\ell$. Specifically,
\begin{equation}
\label{eq:n_c}
    n_c = \frac{1}{\ell-2}.
\end{equation}
We are also able to exactly identify the correlation length exponent $\nu = 2$, which is universal to all models of this type. Numerical simulations suggest a %n unusually 
large dynamical critical exponent $z_c\gtrsim 6$, consistent with the results in Ref.~\cite{PhysRevB.101.214205}.

\section{Model}
\label{sec:model}

\begin{figure}[tb!]
\begin{center}
\includegraphics[width=0.9\columnwidth]{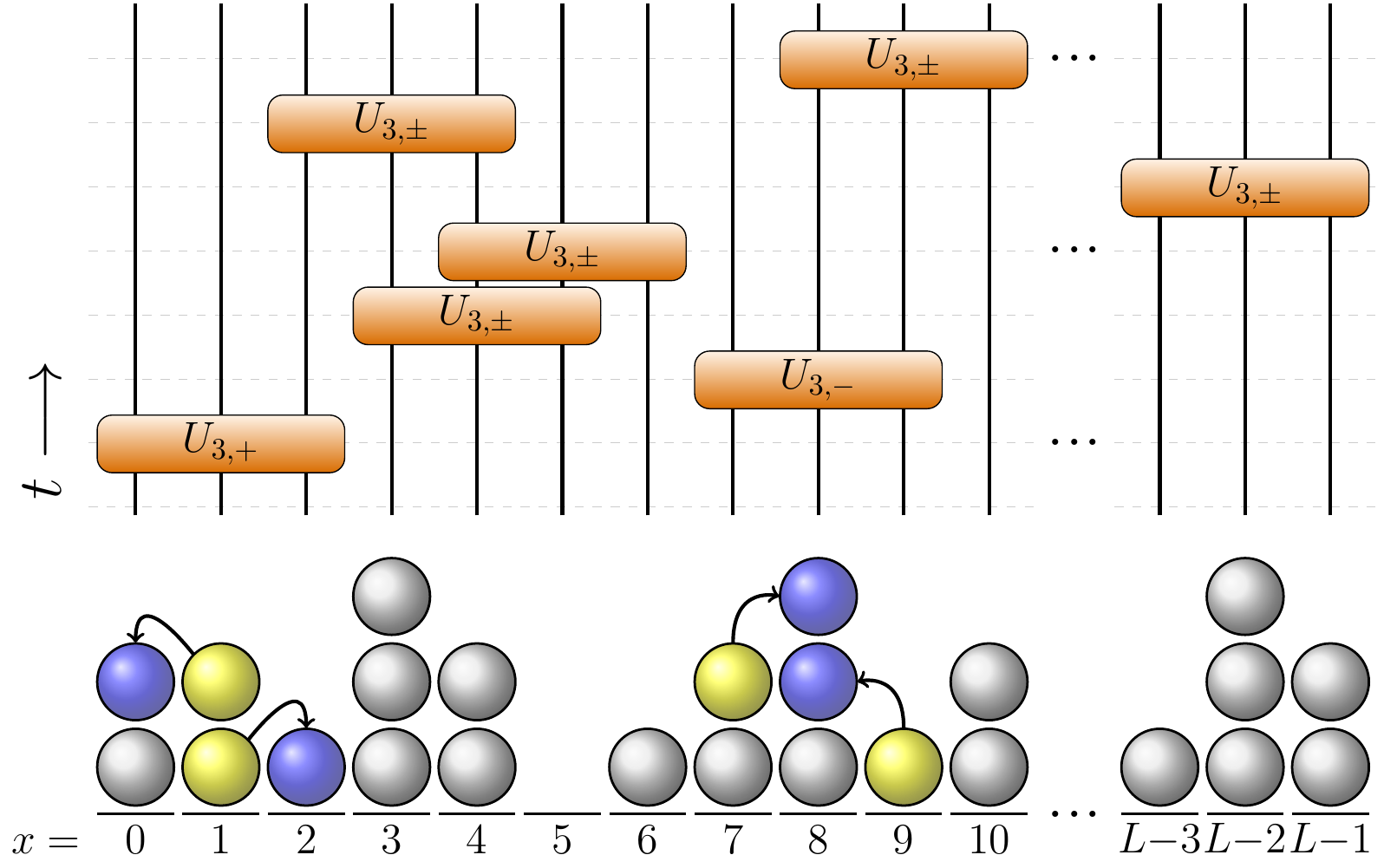}
\end{center}
    \caption{An illustration of the dynamics with charge- and dipole-conserving 3-site gates $U_{3,\pm}$. The circuit (above) shows the sequence of random operations, while the balls (below) illustrate the occupation numbers of the state. Yellow balls represent the starting positions of particles involved in the first two applied operators, blue balls represent the final positions of these particles, and grey balls show particles that remain in place. These two operations are the only %nontrivial dipole and charge conserving 
    3-site gates for our model. %Each attempted gate is considered as $\Delta t = 1/L$ to ensure that that $\Delta t = 1$ corresponds to the time it takes for an average of one gate to be applied at each site. 
    }
    \label{fig:localgate}
\end{figure}

We consider a ``bosonic" system of $N$ indistinguishable particles moving on a 1-D lattice of size $L$ with closed boundary conditions. Each site  $x$, with $x=0,1,...,(L-1)$, has an occupation number given by a non-negative integer, {$n_x = 0, 1, 2, \ldots$} . The system evolves by a random sequence of $\ell$-site local gates, as illustrated in Fig.~\ref{fig:localgate}. The gates are each chosen randomly from the set of operators that conserve both the charge $N$ and the dipole moment $P$, with $N = \sum_x n_x$ and $P = \sum_x n_x x$. %Here $x_i$ is the position of site $i$, which we define such that $x_i = 0$ at the left boundary.
Since the fragmentation of the Hilbert space arises from the classical charge and dipole moment constraints, we are able to restrict our attention to an effectively classical Markov dynamics for which each operator takes a given  charge state (a string of definite values of $n_x$) to another given charge state. This approach is equivalent to the the recently-described ``automaton dynamics'' method \cite{PhysRevB.100.214301, Alba_Dubail_Medenjak_2019, Iaconis_automata_2021, Gopalakrishnan_automata_2018}.

This restriction of the dynamics to  classical charge states implies that each operator is chosen from a small, finite set. For example, in the case $\ell = 3$ any allowed operator %applied to the three contiguous sites ${i-1, i, i+1}$ 
is a multiple of only two nontrivial actions, which we denote $U_{3, \pm}$. Specifically, $U_{3, \pm}$ makes the transformation $\{ n_{x-1}, n_x, n_{x+1} \} \rightarrow \{ n_{x-1} \pm 1, n_{x} \mp 2, n_{x+1} \pm 1 \}$ for some location $x$, as illustrated in Fig.~\ref{fig:localgate}. In our dynamics, each operator is chosen randomly from one of these possibilities and then applied to a random set of $\ell$ contiguous sites. If the operator does not produce a valid basis state -- i.e., if one of the occupation numbers would become negative -- then no operation is applied.

%Each defining fracton characteristic in this system is the conservation of the dipole moment, $P = \sum_{i=0}^{L-1}n_i x_i$, where $n_i \ge 0$ is the number of particles at position $x_i$. In order to evolve the system, we restrict these particles to moving through the operation of $\ell$-site local gates that conserve total particle number and charge. We begin with the case of $\ell=3$ in which the only allowed local gates are $U_{3,\pm,i}= \{n_{i-1},n_i,n_{i+1}\} = \{\pm 1, \mp 2, \pm 1 \}$ (Fig.~\ref{fig:localgate}). At each time step, a site, $x_i \in \{1,...,L-2\}$ is chosen at random and then either $U_{3,+}$ or $U_{3,-}$ is applied at that position. If the chosen operation is allowed, i.e.~the resulting configuration has $n_i>=0$ for all $x_i$, then the operation is performed. Otherwise, the system remains unchanged in that time step.

%With these dynamics in mind, we attempt to quantify the fragmentation of the Hilbert space in order to measure the critical density, $n_c=N/L$ at which the system thermalizes. Demonstratively, when $n=N/L \ll 1$, the system is comprised of many isolated particles which cannot move while conserving dipole. In the opposite limit of $n \gg 1$, there are many particles per site and thus the probability of an attempted gate not being allowed is vanishingly small. We can quantify this heuristic by exploring the fragmentation of the Hilbert space.

For a given charge $N$ and system size $L$, there is some finite number of basis states which all have the same given dipole moment $P$. We refer to this set of states as the \emph{symmetry sector}. Within the symmetry sector, there may be states which cannot be evolved into one another through the application of only local dipole-conserving gates of size $\ell$. For example, in the case $L=5$, $N = 3$, and $P = 6$, the two states $(1,0,1,0,1)$ and $(0,0,3,0,0)$ are dynamically disconnected when $\ell = 3$ despite belonging to the same symmetry sector. We refer to each subset of the symmetry sector for which any pair of states within the subset can be reached one from another through local gates as a \emph{Krylov sector}. The Krylov sectors are dependent on the gate size $\ell$, while the symmetry sectors are not. For example, when $\ell = 2$ all Krylov sectors contain only a single state, since there are no nontrivial 2-site operators that conserve both charge and dipole moment; in the limit $\ell = L$ each symmetry sector consists of only a single large Krylov sector, since all possible $N$- and $P$-conserving transformations are possible.

With these definitions, we can concretely define a thermalized system in terms of the proportion of states within a symmetry sector that belong to its \emph{largest} Krylov sector (LKS). Specifically, we define the quantity $D = D_\text{LKS}/D_\text{sym}$, where $D_\text{LKS}$ is the number of basis states within the largest Krylov sector and $D_\text{sym}$ is the number of basis states in the corresponding symmetry sector. The thermalized phase is characterized by $D \rightarrow 1$ in the limit $L \rightarrow \infty$ with $n=N/L$ held fixed, while the localized phase exhibits instead $D \rightarrow 0$. %in the limit $L \rightarrow \infty$.

%\section{Results}
%\label{sec:results}

%\subsection{Enumeration of states}
%\label{sec:enumeration}

%\section{The size of the symmetry sector and %largest 
%a given Krylov sector}

\section{Solution for the critical filling}

In this section we present results for the size $D_\text{sym}$ of the symmetry sector and the size  $D_\text{KS}$ of a specific Krylov sector (which, as we discuss below, is apparently the largest Krylov sector). By considering the scaling of $D_\text{sym}$ and $D_\text{KS}$ with the system size $L$, we are able to precisely identify the critical density $n_c$ associated with the thermalization transition.
We restrict our attention primarily to the symmetry sector with dipole moment $P = N(L-1)/2$, whose average local charge density is symmetric about %such that the average position of the fractons is at 
the center of the system. Throughout this section we focus on the smallest nontrivial gate size, $\ell = 3$; the generalization to larger $\ell$ is provided in Sec.~\ref{sec:PBC}.
%as a function of $L$ at $n = 1$. 

\subsection{Scaling conjecture for localized and thermalized regimes}
\label{sec:scaling_conjecture}

We begin by conjecturing that in the localized phase, $n<n_c$, the relative size $D$ of the LKS is exponentially small in the system size $L$, while $1-D$ is exponentially small in the thermalizing phase,  $n>n_c$. This conjecture is supported by numerical observations in Refs.~\onlinecite{khemani_shattering_2020,  Sala2020}, as well as our own numeric simulations. % exact enumerations at small system size ($L<15$) where $D$ can be calculated exactly. 
Under this conjecture, \emph{all} Krylov sectors must occupy an exponentially small fraction of the symmetry sector in the localized phase. Likewise, in the thermalized phase, all Krylov sectors other than the LKS occupy an exponentially small fraction of the symmetry sector.

In the remainder of this section we demonstrate the existence of a particular Krylov sector that occupies a power-law fraction of the symmetry sector at the filling $n=1$. Given our scaling conjecture about $D$, such a Krylov sector can only exist precisely at the critical filling. Consequently the value of $n_c$ must be equal to $n_c = 1$ (for gate size $\ell = 3$). As we argue below, the Krylov sector we identify is very likely to be the LKS, which allows us to study the critical scaling of $D$ near the transition.

\subsection{Size of the symmetry sector}
\label{sec:symmetry}

\begin{figure}[tb!]
\begin{center}
\includegraphics[width=0.8\columnwidth]{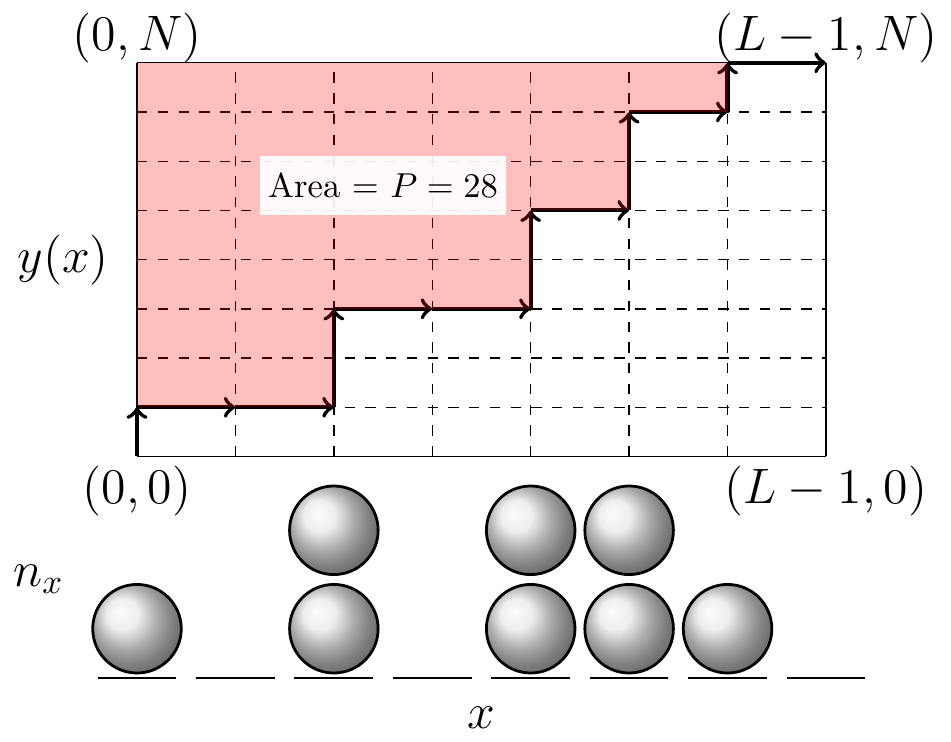}
\end{center}
    \caption{Analogy between non-decreasing integer lattice paths and symmetry sector states. The $x$-axis corresponds to position and each particle corresponds to a one unit move in the $y$-direction. This construction guarantees that the area bounded between the curve and the $y$-axis is equal to the dipole moment of the state.}
    \label{fig:lattice_paths}
\end{figure}

In order to identify the critical filling, we first study how the size of the symmetry sector scales with $L$ and $n$. For this question we can exploit an exact analogy between the number of states in the symmetry sector and the number of non-decreasing lattice paths in a square grid that enclose a fixed area. The key idea is that one can define a ``height field'' $y(x)$ defined for discrete values $x$ by $y(x) = \sum_{w \le x} n_w$  \cite{Feng_Hilbert_2022, Moudgalya_Prem_Huse_Chan2021}. This height field has an endpoint %endpoints $y(0) = 0$ and 
$y(L-1) = N$ that is fixed by the total charge, and an area under the curve $\sum_x y(x) = N(L-1)-P$ that is fixed by the dipole moment. Thus the number of states in the symmetry sector is equal to the number of such curves with fixed endpoint and fixed area. An example is shown in Fig.~\ref{fig:lattice_paths}.

Fortunately, this latter problem has been studied in the mathematical literature \cite{lattice_paths, dobrushin1996fluctuations, perfilev_excursion_2018}. In the limit of large $N$ and $L$, the number of non-decreasing integer lattice paths has been shown to follow \cite{lattice_paths}
\begin{alignat}{2}
    \label{eq:symmetry_sector_asym}
    &D_\text{sym}(N,L,P) \nonumber \\ 
    &\simeq \binom{N+L-1}{N}&&\mathcal{N}\left(P;\frac{N(L-1)}{2},\frac{N(L-1)(N+L)}{12}\right) \nonumber \\
    &\simeq \frac{\sqrt{3}}{\pi n(n+1)L^2} &&\left(\frac{(n+1)^{(n+1)}}{n^n}\right)^L  \nonumber \\
    & &&\times \exp \left[\frac{-6 \tilde{P}^2}{n(n+1)L^2(L-1)} \right].
\end{alignat}
Here $\mathcal{N}(v; \mu, \sigma^2)$ denotes a normal distribution for the variable $v$ with mean $\mu$ and variance $\sigma^2$, and $\tilde{P} = P - N(L-1)/2$ is the dipole moment relative to a coordinate system with its origin at the center of the system. This expression can be roughly understood as follows. If the dipole constraint (or, in analogy, the area constraint) is removed, then the number of lattice paths can be found by straightforward combinatorics to be $\binom{N+L-1}{N}$. Intuitively, symmetric states (lattice paths with area half of the rectangle) are the most likely, and as $L \rightarrow \infty$ the %central limit theorem guarantees that the
likelihood of a given value of $\tilde{P}$ follows a Gaussian distribution.

Notice, in particular, that at $n = 1$ the value of $D_\text{sym}$ at $\tilde{P} = 0$ scales with system size as $ 4^L /L^2 = 4^N / N^2$.

\subsection{Size of the Krylov sector containing the uniform state}
\label{sec:krylov}

Now that the asymptotic scaling of the size $D_\text{sym}$ of the symmetry sector is understood, we consider the fraction of the symmetry sector that is occupied by a specific Krylov sector. In particular, we consider the Krylov sector containing the uniform state ($n_x = 1$ for all $x$). This Krylov sector belongs to the symmetry sector with $N=L$ and $\tilde{P} = 0$. As we now show, for this specific Krylov sector we can make use of another exact analogy to a problem in combinatorics.

\begin{figure}[tb!]
\begin{center}
\includegraphics[width=0.9\columnwidth]{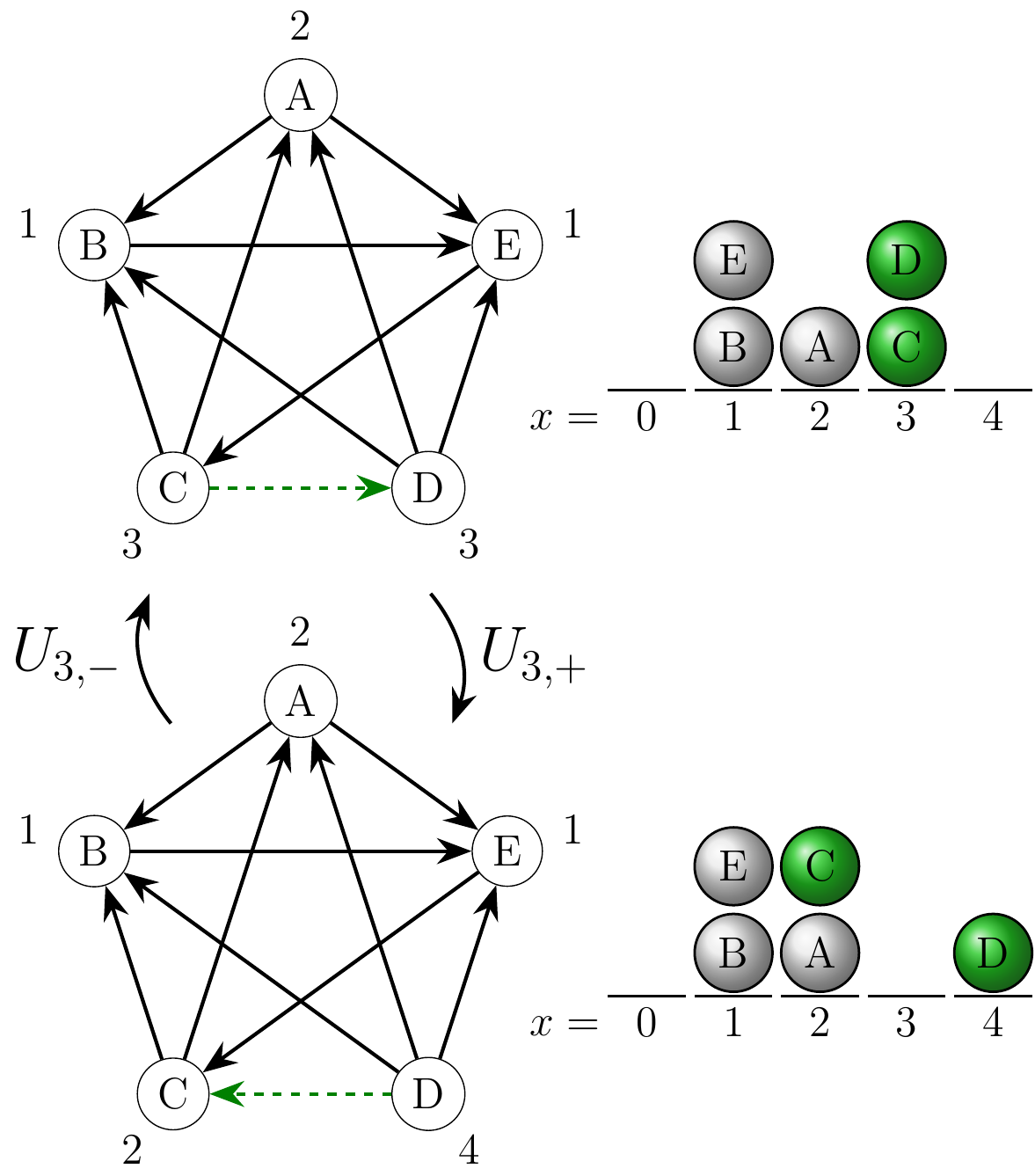}
\end{center}
    \caption{Analogy between tournament scoring sequences and states in the apparent LKS. On the left are tournament graphs for $N = 5$ teams, with the score of each team (the number of outgoing edges) labeled. A given scoring sequence corresponds to a particle distribution, with the number of wins for each team corresponding to the position $x$ of a particle. In this analogy, it is clear that $U_{3,\pm}$ corresponds to flipping the result of a game between two teams that that have either the same number of wins ($U_{3,+}$) or a number of wins that differ by 2 ($U_{3,-}$). Note, particles are indistinguishable in our dynamics and in this figure are only labeled for clarity.}
    \label{fig:tournament}
\end{figure}
%To find such a state we consider the next higher moment of a given configuration, $Q=\sum_{i=0}^{L-1}n_i x_i^2$. While the dipole moment corresponds to the mean position of the particles, this quadropole corresponds to the spread in the positions of the particles. Further, one can show that applying $U_{3,\pm}$ results in a change of $\pm 2$ to $Q$. Therefore, if we consider the configuration in each connected subsector that has maximal variance, then they are defined as the configurations that have no index that $U_{3,+}$ can be applied to.  we see that in order to maximize the number of connected configurations, we want to minimize the charge stuck on the boundary. For the case of $N=L$, this argument suggests that the configuration with one particle at each position belongs to the LKS.

In order to find the size of the Krylov sector containing the uniform state, we draw an analogy to a classic problem in combinatorics: the number of unique scoring sequences of an $N$-team round robin tournament \cite{tournament, conjecture}. A round robin tournament is a directed graph in which each of the $N$ nodes (``teams'') is connected to all $N-1$ other nodes, as shown in Fig.~\ref{fig:tournament}. An outgoing (incoming) edge at a particular node corresponds to that team winning (losing) its match-up with the team at the other end of the edge. An ordered list of numbers of outgoing edges from each vertex makes up a ``scoring sequence'' for a given tournament graph -- that is, a ``scoring sequence'' is the rank-ordered record of how many games were won by each team. We make an analogy to this problem by relating the number of wins by each of the $N$ teams to the positions of the $N$ particles in our system. %The position of a given team corresponds to the position of the relating particle. 

For example, a tournament in which one team loses all of their games, one teams wins one game, one teams wins two games and so on would have a scoring sequence $\{0,1,2,...,N-1\}$. In our analogy, this sequence corresponds to a single particle at each lattice position. Furthermore, the action of a local gate, $U_{3,\pm}$, is analogous to flipping the outcomes of certain games in the tournament. If two teams have the same number $x$ of wins (by analogy, two particles have the same position $x$) and then the result of the game between them is flipped, then one team decreases its win total by $1$ and the other team increases its win total by $1$ (one particle hops left to position $x-1$ and one hops right to position $x+1$). Thus, by flipping the outcome of this game we have performed a $U_{3,+}$ gate centered at the position $x$. Similarly, by flipping the outcome of a game in which a team with $x+1$ wins defeated a team with $x-1$ wins, we can effectively apply a $U_{3,-}$ gate. While flipping the outcome of some games would effectively implement longer range gates (if the two teams involved have a number of wins that is different by more than $2$), we show in Appendix~\ref{app:tournament_proof} that there is a one-to-one mapping between the set of states within  this Krylov sector and the set of scoring sequences, so that the effect of any such long-ranged gate can be equivalently produced by a sequence of local gates.  Thus, we have shown that the number of states in the Krylov sector that contains the uniform state is equivalent to the number of unique scoring sequences in an $N$-team round robin tournament.

Having made this analogy, we can understand the size of this Krylov sector by looking up the result for the number of unique scoring sequences in the mathematical literature. Specifically, Refs.~\onlinecite{tournament, conjecture} show that
\begin{equation}
    \label{eq:D_Krylov}
    D_\text{KS} \sim \frac{4^{N}}{N^{5/2}}
\end{equation}
at large $N \gg 1$. 

One can now compare Eq.~(\ref{eq:D_Krylov}) with the size $D_\text{sym}$ of the corresponding symmetry sector, given by Eq.~(\ref{eq:symmetry_sector_asym}). For the corresponding density $n=1$ and dipole moment $\tilde{P}=0$, Eq.~(\ref{eq:symmetry_sector_asym}) gives $D_\text{sym} \sim 4^N/N^2$, which means that the Krylov sector containing the uniform state occupies a fraction $D \sim 1/N^{1/2}$ of the symmetry sector.
From the exponential scaling conjecture of Sec.~\ref{sec:scaling_conjecture}, a Krylov sector occuping a power-law fraction of the symmetry sector can only exist precisely at $n=n_c$. Hence we conclude that $n_c = 1$.

\subsection{Size of the LKS at $n \leq n_c$}
\label{sec:asymptotic}

We now conjecture that the Krylov sector containing the uniform state, considered in the previous subsection, is precisely the LKS at $n = 1$ and $\tilde{P}=0$. 
This conjecture can be checked by explicit numerical enumeration of all states in the symmetry sector when $L$ is not too large; this procedure confirms our conjecture for $L \leq 15$.

Another way to motivate the conjecture that the LKS contains the uniform state is to notice that any Krylov sector can be uniquely labeled by a ``fully extended state'' for which no $U_{3,+}$ operations can be applied. Such a fully extended state must have $n_x=0$ or $1$ for all $1\le x\le L-2$, with any remaining charges on the boundaries ($x = 0$ and $x = L-1$).  One can show (see Appendix~\ref{app:unique_configurations}) that no two different states satisfying both of these criteria can belong to the same Krylov sector. By labeling each Krylov sector by its corresponding fully extended state, we can now identify the LKS by selecting the fully extended state that allows for the largest number of other states to be accessed through successive applications of the $U_{3,-}$ operation. Since the interior of a fully extended state is sparsely populated, excess particles on the boundary are effectively trapped for $N=L$ and cannot spread into the bulk of the system. Therefore, states with many charges on the boundary are dynamically connected to relatively few other states. This intuition suggests that the fully extended state corresponding to the LKS is the one with the least amount of charge on the boundaries. For the case of $N=L$, this state is precisely the uniform one. Hence the relative size of the LKS $D(n=1) \sim 1/\sqrt{L}$.

Let us now extend this result to the case $n < 1$, for which $N < L$. The key idea is that for $n < 1$ the LKS still contains a fully extended state with a long chain of $\sim N$ successive $1$'s, surrounded by zeros on either side. Applying gates to this state may change the occupation numbers in the middle of the chain, but the surrounding zeros always remain inert. Thus, the corresponding LKS is very similar to that of a system at the critical filling and a smaller system size $n L$. 

More precisely, we can argue, using similar logic as above, that the LKS must either contain the state ${A =\{0,...,0,1,0,1,1,1,...,1,1,1,0,1,0,...0\}}$ if $N$ and $L$ have the same parity (both odd or both even), or the state ${B =\{0,...,0,1,1,...1,1,0,1,1...,1,1,0,...0\}}$ if $N$ and $L$ have opposite parity. $A$ and $B$ represent ``nearly-uniform'' states in the case where the number of particles is not large enough to fill the entire system uniformly. Here we focus on the case of $N$ and $L$ having the same parity, although the reasoning for both cases is the same. We now compare the number of states that are accessible starting from state $A$ to the number of states that are accessible from a state with $N-2$ centered particles (i.e., removing the leftmost and rightmost particles from state $A$) and to the number of states that are accessible from a state with $N+2$ centered particles (i.e., adding two new particles to fill the empty spaces just to the right of the leftmost particle and just to the left of the rightmost particle in state $A$). Let us refer to the size of the Krylov sectors containing these two modified states as $D_{N-2}$ and $D_{N+2}$, respectively. The number that we care about, $D_\textrm{LKS}$, is bounded from below by $D_{N-2}$ and from above by $D_{N+2}$. Since $D_{N\pm2}$ describe the number of LKS states in a system of size $N \pm 2$ with uniform filling $n_c = 1$, it follows that ${ 4^{(N-2)}/(N-2)^{5/2} \le D_\text{LKS}\le  4^{(N+2)}/(N+2)^{5/2} }$. From this inequality we conclude that
\begin{equation}
    \label{eq:krylov_n<1}
    D_\text{LKS} \sim \frac{4^{nL}}{(nL)^{5/2}}
\end{equation}
for all $n\le 1$.

Equations (\ref{eq:symmetry_sector_asym}) and (\ref{eq:krylov_n<1}) allow us to write down the relative size $D$ of the LKS for $n \leq 1$ for the case of a symmetric dipole moment ($\tilde{P} = 0$):
\begin{equation}
    \label{eq:D}
    D = \frac{D_\text{LKS}}{D_\text{sym}}\sim \frac{n+1}{n^{3/2}\sqrt{L}}\left(\frac{(4n)^n}{(n+1)^{(n+1)}}\right)^L.
\end{equation}
The second equality corresponds to the limit $nL \gg 1$.

Notice that the factor in parentheses, ${ (4n)^n/(n+1)^{(n+1)} }$, is smaller than unity for all $n<1$, and thus the LKS (and subsequently all other subsectors) occupies an exponentially small portion of the symmetry sector as $L \rightarrow \infty$. This exponential scaling is consistent with our conjecture in Sec.~\ref{sec:scaling_conjecture}, and demonstrates that $n<1$ corresponds to strong fragmentation. Exactly at the critical filling, the relative size of the LKS has a power-law decay with system size:
\be 
D(n=n_c) \sim \frac{1}{\sqrt{L}}.
\label{eq:Dnc}
\ee 
Equations (\ref{eq:D}) and (\ref{eq:Dnc}) are verified numerically in Fig.~\ref{fig:powerlaw}. %Further, our conjecture regarding the form of the LKS for $n<1$ implies our previous conjecture that all Krylov sector be exponentially small in the localized phase. 

In Appendix \ref{app:general_ell} we generalize the argument in this section to arbitrary gate size $\ell \ge 3$, and we  obtain $n_c = 1/(\ell-2)$. In Sec.~\ref{sec:PBC} we present an alternative, shorter derivation of this result for $n_c$ by considering a system with periodic boundary conditions.

%\begin{figure}[tb!]
%\begin{center}
%\includegraphics[width=0.9\columnwidth]{Fig_1A.pdf}
%\includegraphics[width=0.45\columnwidth]{Fig_1B.pdf}
%\includegraphics[width=0.45\columnwidth]{Fig_1C.pdf}
%\includegraphics[width=0.45\columnwidth]{Fig_1D.pdf}
%\end{center}
    %\caption{The proportion $D$ of states that belong to the largest Krylov sector (LKS), as measured by an exact enumeration of states, plotted as a function of the filling $n$. Different curves correspond to different system sizes, ranging from $L = 11$ (lightest curve) to $L = 101$ (darkest curve). The dashed line shows the behavior in the limit $L \rightarrow \infty$.
    %(B) $D$ scaled according to Eq.~\ref{eq:D} with the numeric prefactor taken to be unity. (C) $D$ at the critical filling $n = n_c = 1$ decays as $D\sim L^{-1/2}$ (D) For periodic boundary conditions, the system is ergodic for $N = L-1$ and strongly fragmented for $N=L-2$. As $L \rightarrow \infty$ both cases approach $n=1$, thus implying that $n_c=1$.
    %}
    %\label{fig:enumeration}
%\end{figure}

\subsection{Numerical results for $D$}

\begin{figure}
\begin{center}
\includegraphics[width=0.9\columnwidth]{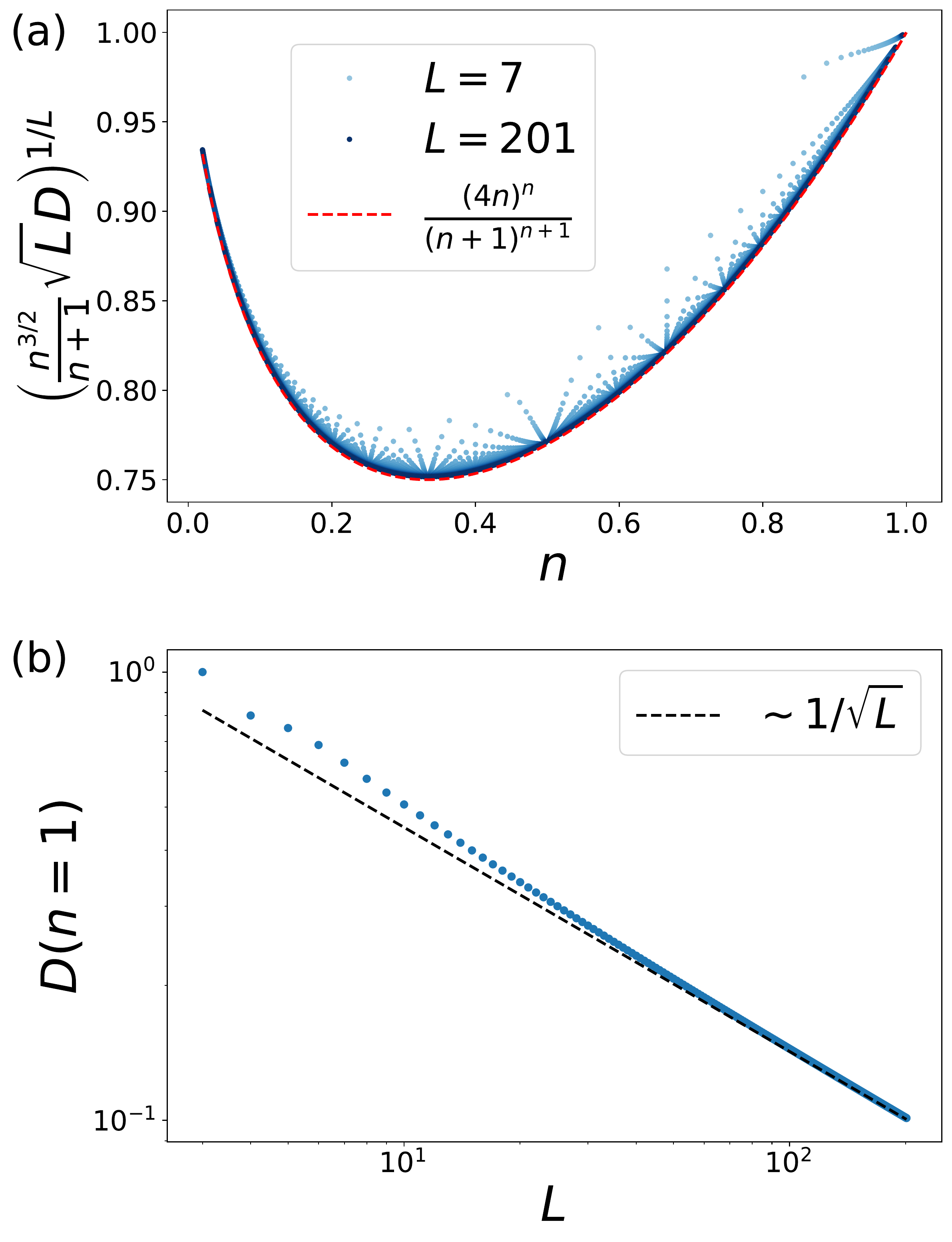}
\end{center}
    \caption{The relative size of the apparent LKS as a function of filling $n$ and system size $L$. (a) For $n\le 1$, the relative size $D$ of the LKS approaches the scaling suggested by Eq.~\ref{eq:D} in the limit $L\rightarrow \infty$ with no fitting parameters. Increasingly dark symbols correspond to progressively larger system size. (b) Exactly at the critical filling, $n=1$, $D$ decays as $\sim L^{-1/2}$. This transition from exponential decay at $n<1$ to power law decay at $n = 1$ is indicative of the thermalization transition.}
    \label{fig:powerlaw}
\end{figure}

The most straightforward numerical procedure for studying the relative size, $D$, of the LKS is to list all the possible states that have a given $N$ and $P$ and then sort them into their respective Krylov sectors for $\ell=3$. This type of exact enumeration is only possible for sufficiently small $L$ and $N$, since the size of the symmetry sector grows exponentially in $L$. However, recursive algorithms based on the analogy presented in the previous subsections, and detailed in Appendix~\ref{app:algo}, can be used to extend to larger system sizes. In Fig.~\ref{fig:powerlaw}(a) we verify the scaling of $D$ (for the conjectured LKS) given by Eq.~(\ref{eq:D}) at $n<1$. Figure~\ref{fig:powerlaw}(b) shows the value of $D$ at $n = n_c =1$, which verifies Eq.~(\ref{eq:Dnc}).

\section{Extension to arbitrary gate size}
\label{sec:PBC}

\begin{figure*}
\begin{center}

\includegraphics[width=0.9\textwidth]{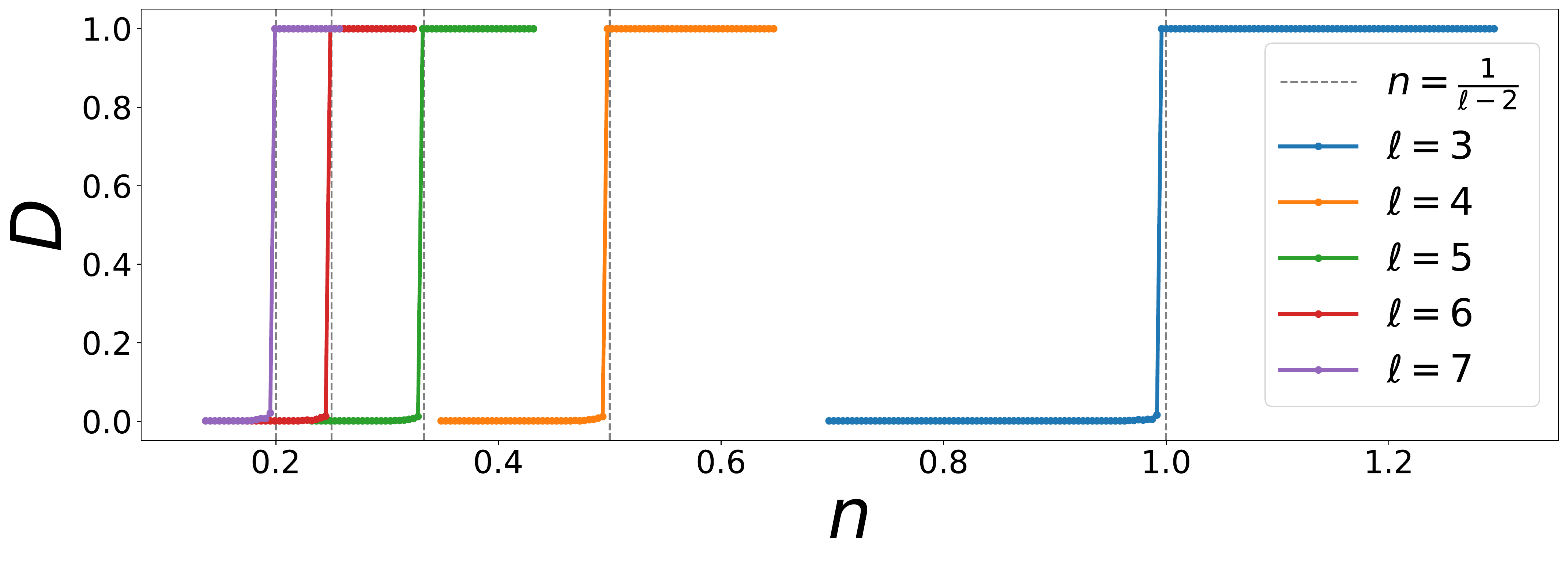}
\end{center}
    \caption{The relative size $D$ of the LKS for systems with periodic boundary conditions, as measured by numeric simulations. The value of $D$ is plotted as a function of the filling $n$ for different values of the gate size $\ell$. Data shown here corresponds to system size $L = 241$ and is averaged over 1000 random choices of the initial state (see the Appendix \ref{app:random_state} for a full description of the simulation protocol). The vertical dashed lines show the predicted critical filling $1/(\ell - 2)$.
    }
    \label{fig:periodic}
\end{figure*}

So far we have focused primarily on the case of dynamics with 3-site gates. We now consider the extension of our results to arbitrary (integer) gate size $\ell \geq 3$. Our goal is to demonstrate that the critical density $n_c = 1/(\ell - 2)$, as mentioned in the introduction.

Our strategy for proving that $n_c = 1/(\ell - 2)$ is to consider the case of two slightly different values of $n$ that both approach $1/(\ell - 2)$ in the limit $L \rightarrow 0$. We show that the larger of these two fillings has $D \rightarrow 1$ in the limit $L \rightarrow \infty$, while the smaller filling gives $D \rightarrow 0$ in the limit $L \rightarrow \infty$. This difference establishes the critical filling $n_c = 1/(\ell -2)$ in the limit $L \rightarrow \infty$. 

Our arguments are considerably simplified by focusing on the case with periodic boundary conditions, which produces the same critical density in the limit $L \rightarrow \infty$. In this case, we should be careful to define the dipole moment modulo the system size, so that its value is unchanged when, say, two particles are initially at $x=0$ and then one hops to ${x = L - 1}$ while the other hops to ${x = 1}$. Thus for periodic boundary conditions we define the dipole moment as ${P = \left(\sum_{x=0}^{L-1} n_x x\right)\text{ mod }L}$.
If we further restrict our consideration to values of $N$ and $L$ that are coprime, then we can show that all possible values of $P$ produce equivalent sets of states and are therefore equivalent to each other. This equivalence is apparent if one imagines the process of taking a particular state and shifting the origin of the coordinate axis $x$. This shift produces an equivalent state with a different value of $P$. By doing such shifts one can reach any value of $P$, and therefore when $N$ and $L$ are coprime all possible values of $P$ have the same $D_\text{sym}$ and $D_\text{LKS}$.

%We now turn our attention to studying the system with periodic boundary conditions (which should not change any of the physics in the thermodynamic limit) and generalize to any $\ell$-site local gate that conserves charge and dipole moment. Here, we must be careful to define the dipole moment in a manner that avoids the dipole moment changing as a particle moves from the $x_i=L-1$ site to $x_i=0$. Thus, when the boundaries are periodic, the dipole moment must be defined as $P = \left(\sum_{i=0}^{L-1} n_i x_i\right)\text{ }(\text{mod }L)$. Further, if we limit ourselves to considering parameters where $N$ and $L$ are coprime, then shifting our system by one site, simply changes our dipole. Therefore, for a given coprime $N$ and $L$, the symmetry sectors for each $P$ has the same number of configurations, $\binom{N+L-1}{N}/L$. Therefore, we need not worry about what $P$ we choose since all are identical up to translation.

It is also useful to note that any local $\ell$-site gate that conserves both charge and dipole moment can be decomposed into a sum of simple gates that correspond to pairs of particles hopping either toward or away from each other by one site. That is, we need only consider the successive application of gates $U_{k,\pm} = \{\pm 1, \mp 1,0,...,0,\mp1, \pm 1 \}$, where $3 \leq k \leq \ell$ and there are $k - 4$ zeros.

We begin by considering a system with $N$ particles and a number of lattice sites given by $L = N(\ell - 2) + 1$, so that the average particle density  $n = [1 - 1/L]/(\ell - 2)$ is slightly smaller than $1/(\ell - 2)$.  We now imagine the process of producing a ``fully extended state'' starting from an arbitrary initial state within a particular symmetry sector (we need not specify the value of $P$ since for the case of periodic boundaries all values of $P$ are equivalent).  Specifically, we repeatedly apply $U_{k,+}$ operators (for $3 \le k \le \ell$) until there is at most one particle in every set of $\ell -2$ contiguous sites. (Note that if any set of $\ell - 2$ contiguous sites has more than one particle, then there is some operator $U_{k,+}$ that can be applied.) This procedure can only produce a single unique final state, comprising $N$ units of the sequence $\{1, 0, ..., 0\}$ with $\ell - 3$ zeros, and one additional zero whose position determines the dipole moment $P$. Thus, since any arbitrary initial state can be connected to the same fully extended state, it follows that all states within the symmetry sector belong to the same Krylov sector, and hence that $D = 1$. (Notice that for the case of periodic boundary conditions we have $D = 1$ exactly at $n \geq n_c$, even for finite $L$, unlike the case of closed boundary conditions.) 
In Appendix \ref{app:fully_extended_PBC} we demonstrate more rigorously that all states with $L=N(\ell-2)+1$ are dynamically connected to a unique fully extended state.
%\calvin{We demonstrate all configurations must reach a unique fully extended state in Appendix \ref{app:fully_extended_PBC}.}

%Through a recursive argument we can now show that $D$ must remain one if more particles are added. Imagine taking the symmetry sectors of $P=0,1,...L-1$ and adding an additional particle to position $x_i = P'-P \text{ }(\text{mod } L)$. Through this method, we are guaranteed that all of the symmetry sectors now have the same dipole moment $P'$ and are individually connected. Additionally, there must be overlap between each of these previous symmetry sectors. Therefore, all of the new states that have $N+1$ particles and dipole moment $P'$ have $D=1$. This argument can be repeated for any number of additional particles show that $D=1$ for $n > 1/(\ell-2)$.

Let us now consider the process of constructing a fully-extended state from an initial state with one additional lattice site, $L = N(\ell-2)+2$, so that $n = [1 - 2/L]/(\ell - 2)$ is slightly smaller than in the previous case. Repeated applications of $U_{k,+}$ eventually produce a fully extended state that is similarly composed of many repeating units $\{1, 0, ..., 0\}$ with $\ell - 3$ zeros. There are still $N$ such units, but, unlike in the previous case, there are now two additional zeros to be placed somewhere among them. Since there is more than one extra zero to be placed, the positions of these extra zeros are not uniquely specified by the dipole moment $P$. Indeed, the number of possible positions for the zeros in the fully extended state grows linearly with the system size $L$, and different fully extended states cannot be evolved one into another. 
%Each of these fully extended states consists of $L-2$ units of $\{1, 0, ..., 0\}$ with two additional zeros.
Since there is an extensive number of fully extended states, each belonging to a different Krylov sector, it is natural to conclude that the symmetry sector must become increasingly fragmented as $L \rightarrow \infty$, and consequently that $D \rightarrow 0$.  

Formally, this last logical step has the status of a conjecture: we are conjecturing that none of the $\sim L$ distinct Krylov sectors is dominant in the sense of occupying all but an exponentially small portion of the symmetry sector.  But, given that the difference between the various fully extended states that label the Krylov sectors is only the placement of two zeroes, we consider it to be a very natural conjecture, which implies that $D \rightarrow 0$ in the limit $L \rightarrow \infty$.

Thus, since we have demonstrated that a density $n = [1 - 1/L]/(\ell - 2)$ produces $D = 1$ and a density $n = [1 - 2/L]/(\ell - 2)$ produces $D \rightarrow 0$ in the limit $L \rightarrow \infty$, it follows that the critical density must be equal to $1/(\ell - 2)$.

%We have argued that for $N = (L-1)/(\ell-2)$, $D = 1$ while for $N = (L-1)/(\ell-2) - 1$, $D \rightarrow 0$ in the infinite system size limit. %The case of $\ell = 3$ is shown numerically in Fig.~\ref{fig:periodic}.
%Since we can reach $D=0$ or $1$ using these two different limits of $n\rightarrow 1/(\ell-2)$ as $L\rightarrow \infty$, we have shown that for a dipole conserving fracton system with local $\ell$-site gates, the critical density required for thermalization is exactly given by $n_c = 1/(\ell-2)$.

% I think this paragraph should be moved to the end. 
We numerically confirm the relation $n_c = 1/(\ell - 2)$ using simulations of systems with periodic boundary conditions. Our approach is to begin with a randomly selected initial state from the symmetry sector with a given $P$ and then repeatedly apply $U_{k,+}$ operations (for $3\le k\le \ell$) until the system has reached a fully extended state. We repeat this process for many random choices of the initial state, and we estimate $D$ by the frequency with which the most commonly-encountered fully extended state is reached. The results are shown in Fig.~\ref{fig:periodic} for different gate sizes ranging from $\ell = 3$ to $\ell = 7$. 
%First, we explore the general gate size numerically with periodic boundary conditions in Fig.~\ref{fig:periodic}. To do so, we generate 1000 random configurations from the symmetry sector with given $N,P,$ and $L$. Then, we repeatedly apply $U_{k,+}$ operations, for $3\le k\le \ell$ to reach the fully spread state. Once we do this for all random configurations, we estimate $D$ by taking the number of the most common fully spread state and dividing that by the total number of random configurations. The results (Fig.~\ref{fig:periodic}) show a sharp transition at $1/(\ell-2)$ which we now make an analytic argument for.

\section{Critical exponents of the thermalization transition}

\subsection{Correlation length exponent}
\label{sec:correlation}

Using the previous result for the critical density $n_c$, we can explore the critical behavior near the transition. We first examine the correlation length exponent, $\nu$, defined by $\xi \propto 1/(n_c-n)^\nu$. Here, $\xi$ has the meaning of the typical length of a locally thermalized region within the nonthermalizing phase, $n < n_c$. 
Within such locally thermalized regions, the local particle density exceeds $n_c$. As the global density $n$ is increased towards $n_c$, the typical length of these segments diverges. 

The universal value of the correlation length exponent $\nu = 2$ can be seen by the following simple argument. If the average particle density of the system is $n < n_c$, then a randomly-chosen region of size $L_0 \gg 1$ has a charge $N_0$ that is drawn from a probability distribution with mean $\mu_0 = n L_0$ and a variance $\sigma_0^2$ that is proportional to the number of sites $L_0$ in the region. In order for the region to be locally thermalized, the number of charges in the region should exceed $n_c L_0$. Such a statistical fluctuation is reasonably likely only when $N_0 - \mu_0$ is of order $\sigma_0$ or smaller. Equating these two quantities gives an expression for the typical length $\xi = L_0$ of a locally thermalized region, $(n - n_c) \xi \sim \sqrt{\xi}$, or in other words $\xi \sim 1/(n_c - n)^2$. 

At a more precise level, one can calculate the probability $p(n, L_0; n_c)$ that a region of length $L_0$ contains at least $n_c L_0$ particles. In Appendix \ref{app:nu} we present a full calculation of this probability along with numerical results for $p(n, L_0; n_c)$ obtained by randomly sampling the symmetry sector. We find that the probability $p(n, L_0; n_c)$ decays exponentially at large $L_0$ as $\exp(-L_0 / \xi)$, with
\be 
\xi \simeq \frac{2 n_c (n_c + 1)}{(n_c - n)^2}.
\label{eq:nu}
\ee 
This result establishes that $\nu = 2$.

\subsection{Dynamical Exponent}
\label{sec:dynamical}

Our numerical simulations also enable us to estimate the dynamical exponent $z_c$ that describes the characteristic timescale of the dynamics at the critical point. 
We characterize the dynamics using the time- and position-dependent two-point correlation function
\be 
C(x,t) = \langle (n_{x_0+x}(t_0+t)-n)(n_{x_0}(t_0)-n)\rangle
\ee 
where $n_{x}(t)$ denotes the particle number at site $x$ and time $t$, and $\langle . \rangle$ denotes an average over all choices of $x_0$ and $t_0$. This correlation function $C(x,t)$ can be defined by simulating the dynamics via the circuit in Fig.~\ref{fig:localgate} starting from an initial state that is chosen randomly from the set of all basis states in the symmetry sector. We define our unit of time such that $L$ gates are applied during one time step. Results for $C(x,t)$ are produced by averaging over many random choices of the initial state and its subsequent evolution.

In order to estimate the dynamical exponent, we attempt to scale the position coordinate such that curves $C(x,t)$ corresponding to different times $t$ collapse onto a single curve when plotted as a function of $x/x_0(t)$ for some choice of $x_0(t)$. A natural choice is to define $x_0(t)$ as the position of the first zero of the correlation function at the time $t$, i.e., $C(x_0(t),t)=0$. If we assume that $x_0(t)$ takes the form of $x_0(t) \sim t^{1/z}$, then we can extract the dynamical exponent, $z_c$, from fitting this curve. Performing this fit for early times, $10 < t < 10^5$, gives a value $z_c \approx 5.2$, while doing so at later times,  $10^{5.5} < t < 10^7$, gives $z_c \approx 6.2$. Therefore, we can say that estimating $z_c$ in this way gives $z_c = 5.7 \pm 0.5$.
This result is in qualitative agreement with Ref.~\cite{PhysRevB.101.214205}, which found a slow dynamical exponent of $z_c\gtrsim 7$ in a similar system with constrained dynamics. This large value of $z_c$ should be contrasted with the universal hydrodynamics $x \sim t^{1/4}$ that has been established in the thermalizing phase for dipole-conserving fracton systems \cite{PhysRevResearch.2.033124,PhysRevE.103.022142,PhysRevB.101.214205,PhysRevLett.127.235301,PhysRevE.105.044103, Moudgalya_Prem_Huse_Chan2021}.

We caution, however, that our results do not show convincing scaling of the correlation function; even at the latest times different curves $C(x, t)$ do not completely collapse onto each other when plotted as a function of $x/x_0(t)$. Our numerical results for $x_0(t)$ also show some deviation from the power-law trend at the largest values of $t$, toward (perhaps) larger values of $z_c$.  We thus consider that the dynamics at the critical point remains to be completely understood, and we leave this for future work.

\begin{figure}[tb!]
\begin{center}
\includegraphics[width=0.9\columnwidth]{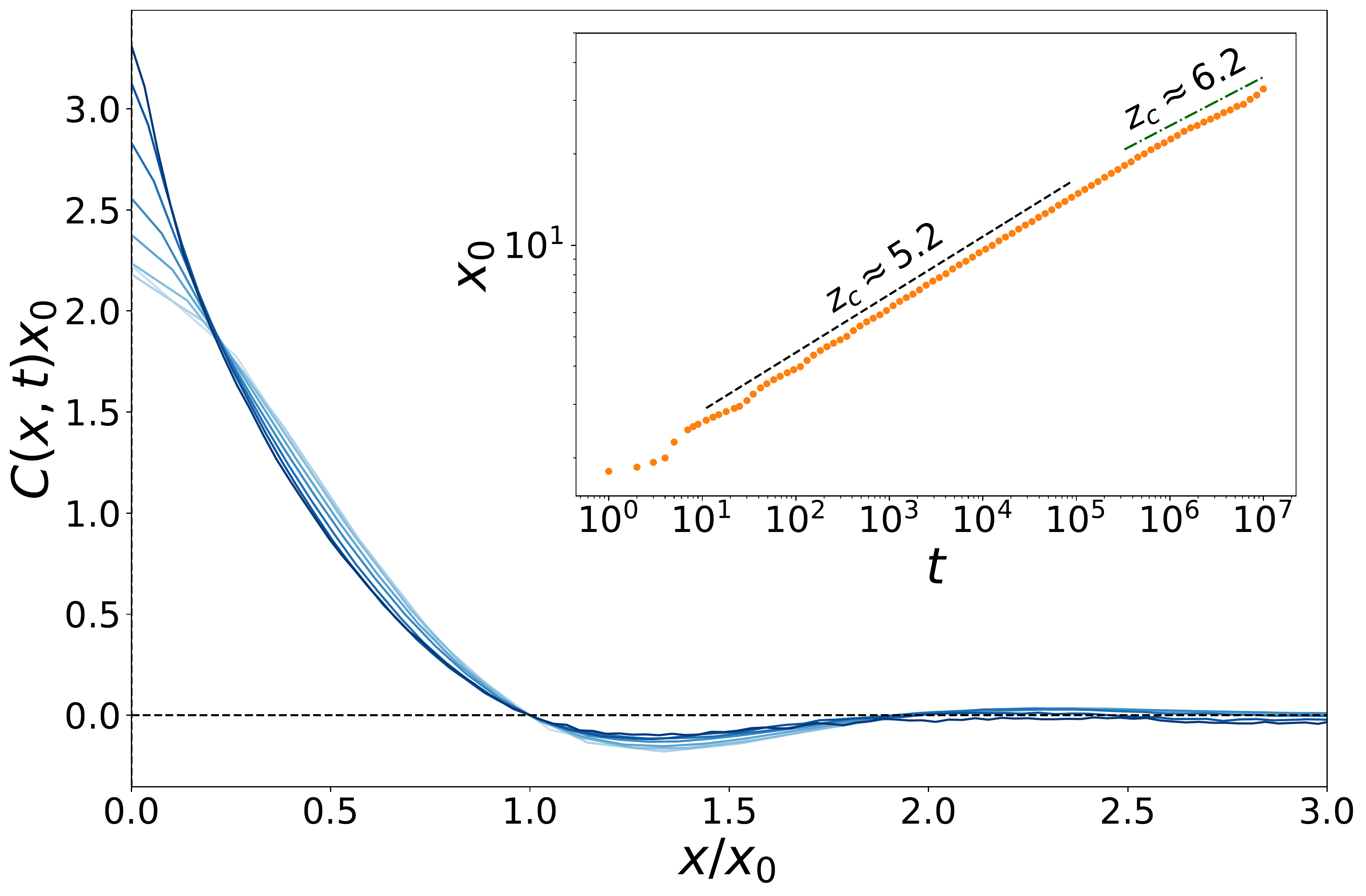}
\end{center}
    \caption{Scaling of the correlation function $C(x,t)$ at the critical point $n = 1$ for dynamics with three-site gates. We scale the position axis by the value $x_0(t)$ at which the correlation function is first equal to zero. Curves correspond to different values of the time, logarithmically spaced from $t = 10^2$ (light blue) to $10^7$ (dark blue). The inset shows that the growth of $x_0$ with $t$ can be fit to a power law with exponent larger than $5$.}
    \label{fig:dynamic}
\end{figure}

\section{Conclusion}
\label{sec:conclusion}

Fracton systems represent a new frontier for the physics of thermalization and localization, in which the thermalization transition is driven by kinetic constraints on the dynamics rather than by any kind of disorder. While such a transition can be effected by discrete variables like the size of local gates or the local Hilbert space dimension at each site \cite{Pai-Nandkishore2019, PhysRevX.10.011047,PhysRevE.103.022142, PhysRevLett.125.245303, Moudgalya_Prem_Huse_Chan2021}, varying the filling $n$ of the system allows one to access the thermalization transition as a continuous phase transition  \cite{PhysRevB.101.214205}. Here we have provided the first exact solutions for the filling-induced thermalization transition, focusing on the critical filling $n_c$ and the dynamical exponent $\nu$. The model we explore differs only slightly from the one in Ref.~\onlinecite{PhysRevB.101.214205}, namely by allowing an unlimited (positive) occupation of each site rather than by restricting each site to occupation numbers $n = 0, 1, 2$. This difference has enabled us to exploit exact analogies to known problems in combinatorics. 
%Through the understanding imported from these analogies, we are able to asymptotically express the degree of ``fragmentation" \cite{khemani_shattering_2020, Rakovszky_Sala_Verresen_Knap_Pollmann2020,  Sala2020, moudgalya_thermalization_2021, Moudgalya_Hilbert_2022}, create recursive algorithms to explore larger system sizes numerically, and explore critical exponents analytically.

It is worth noting that the phenomenology of the transition in our model coincides with what was demonstrated in Ref.~\onlinecite{PhysRevB.101.214205} for the case where the maximal filling at each site $n_\text{max} = 2$, down to the critical filling. In Ref.~\onlinecite{PhysRevB.101.214205}, the authors considered the case of gate size $\ell = 4$ and they found a critical filling which was very close to $1/2 = 1/(\ell - 2)$. (By a ``particle/hole'' symmetry $n \leftrightarrow n_\text{max} - n$, their model also exhibits a transition at $n = 3/2$.) While the analogies we used to derive $n_c = 1/(\ell - 2)$ are not exactly applicable for systems with finite $n_\text{max}$, there may be straightforward arguments to extend our result for the critical filling to such systems.

Also similar to Ref.~\onlinecite{PhysRevB.101.214205}, our model exhibits a large apparent dynamical exponent $z_c$. Our best estimate from scaling of the two-point correlation function gives $z_c \approx 5.7 \pm 0.5$, but given the imperfect scaling and trend toward larger apparent values of $z_c$ at larger times, we take this value to be a lower-bound estimate. Reference~\onlinecite{PhysRevB.101.214205} reports $z_c \approx 7 \pm 0.5$, which they similarly take as a lower bound. Given these large values and the imperfect scaling, it may be that the correlation length at the critical point does not have a power-law scaling with time, and therefore that $z_c$ is not well defined. This conjecture may be a fruitful focus of future work.

\acknowledgments

We are grateful to Alan Morningstar for helpful discussions. %, and to the On-Line Encyclopedia of Integer Sequences for providing a useful connection between the Krylov sector and tournament scoring sequences. 
This work was primarily supported by the Center for Emergent Materials, an NSF-funded MRSEC, under Grant No.~DMR-2011876.  D.A.H. was supported in part by NSF QLCI grant OMA-2120757.

\appendix

\section{Proof of the analogy between the number of states reachable from the uniform state and the number of tournament scoring sequences}

\begin{figure}[tb!]
\begin{center}
\includegraphics[width=0.95\columnwidth]{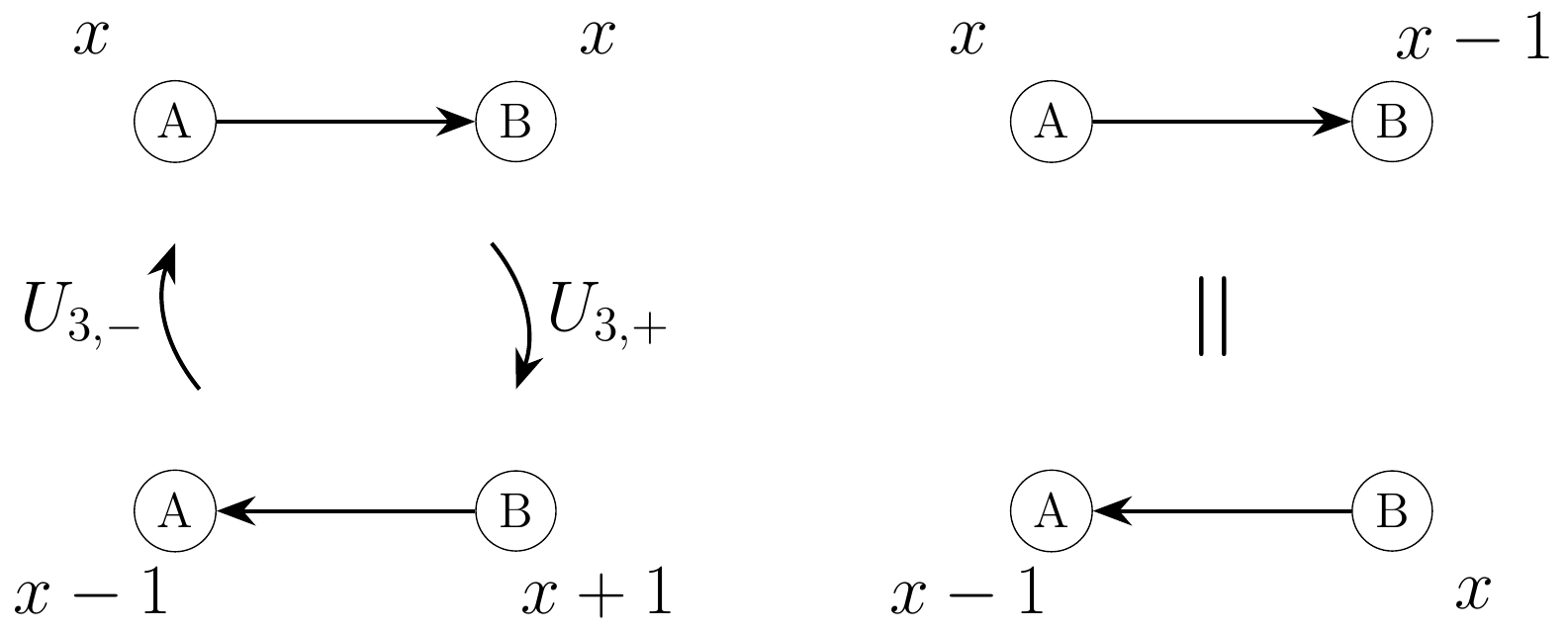}
\end{center}
    \caption{Allowed game flips that correspond to 3-site dipole conserving operations. The game flips on the left correspond to $U_{3,\pm}$ and the game flips on the right leave the scoring sequence unchanged and thus correspond to the identity operation.}
    \label{fig:allowed_flips}
\end{figure}

In Sec.~\ref{sec:krylov} we drew an analogy between the size of the Krylov sector containing the uniform state (conjectured to be the LKS) and the number of scoring sequences in a round robin tournament. In that argument, we demonstrated that flipping the result of a game is analogous to applying a dipole-conserving gate. However, this gate only acts on $\ell=3$ contiguous sites if we flip the result of a game between two teams with the same number of wins (or the reverse of this operation), as shown in Fig.~\ref{fig:allowed_flips}. On the other hand, flipping the result of a game between, say, a team that has 1 win and a team that has 7 wins would be equivalent to a 7-site gate. In this appendix, however, we prove that all scoring sequences can be reached through only 3-site operations.

\emph{Claim:}

A) All tournaments contain the scoring sequence $\{0,1,2,...,N-1\}$. This sequence corresponds to the state with one particle at every site, which we have argued belongs to the LKS.

B) Given a valid scoring sequence $\{x_1,x_2,...,x_N\}$ for an $N$-team round robin tournament, if there are two teams $i$ and $i+1$ such that $x_i = x_{i+1}$, one can apply $U_{3,+}$ and create a new valid scoring sequence. If there are two teams $i$ and $i+m$ such that $x_i+2 = x_{i+m}$, one can apply $U_{3,-}$ and create a new valid scoring sequence. This establishes that all states in the LKS correspond to valid scoring sequences.

C) Starting from any valid scoring sequence for an $N$-round robin tournament, we can reach the scoring sequence $\{0,1,2,...,N-1\}$ through the repeated application of $U_{3,+}$ operators. Therefore, all valid scoring sequences correspond to a state in the LKS.

Together, (A)-(C) show that all valid scoring sequences correspond to a state in the LKS and that all states in the LKS correspond to a valid scoring sequence, thus establishing a one to one mapping between the sets. This mapping guarantees that the sets are the same size.

\emph{Proof:}

A) For $N=1$, the only scoring sequence is $\{0\}$. Assume that $\{0,1,2,...,N-1\}$ is valid for an $N$ team tournament. Then if we add an additional team that beats every other team, the newly added team will have $N$ wins and no other team's score will change. Thus, $\{0,1,2,...,N-1,N\}$ is a valid scoring sequence for an $N+1$ tournament. By induction, part (A) is proven.

B) 
Intuitively, $U_{3,\pm}$ corresponds to flipping the outcome of a game with certain conditions, as shown in Fig.~\ref{fig:allowed_flips}, and therefore still produces a valid tournament. More formally, a valid scoring sequence $\{x_1,x_2,...,x_{N}\}$ is defined by three criteria \cite{landau}:
\begin{enumerate}
\item $0 \le x_1 \le x_2 \le ... \le x_{N} \le N-1$
\item $y_k = \sum_{i=1}^{k} x_i \ge \binom{k}{2}$.
\item $y_{N} = \sum_{i=1}^{N} x_i = \binom{N}{2}$.
\end{enumerate}

\label{app:tournament_proof}
We will show that the action of a 3-site gate preserves these three conditions. Firstly, condition $(3)$ is always satisfied under the application of $U_{3,\pm}$ since both simply shift a win from one team to another while the total, $y_N$, remains fixed.

Now, let us assume that there are exactly $m \ge 2$ teams with the same score, so as to allow for the application of $U_{3,+}$. By requirement $(1)$, these teams are consecutive in the scoring sequence and thus $x_{i} < x_{i+1} = x_{i+2} = ... = x_{i+m} < x_{i+m+1}$. Once the $U_{3,+}$ gate is applied, $x_{i+1}$ will decrease by one while $x_{i+m}$ increases by one, leaving the order the same. Consequently, $y_{i+1}$ through $y_{i+m-1}$ will all decrease by one while the remaining $y_i$'s remain the same. Therefore, in order for this operation to be legal, we must require that the original $y_{i+1}$ through $y_{i+m-1}$ satisfy requirement $(2)$ with a strict inequality. In order to show that this strict inequality is always obeyed, we will consider by contradiction the case when condition $(2)$ is an equality for $y_k$ with $i+1\le k \le i+m-1$ (i.e.~$\exists y_k$ such that $y_k = \binom{k}{2}$). One can see the intuitive meaning of this equality condition by noting that any subset of $k$ teams in a tournament plays exactly $\binom{k}{2}$ games amongst themselves. Therefore, since all of these $\binom{k}{2}$ games appear as wins for teams in that subset condition $(2)$ is an equality when those $k$ teams lose all games against teams not in the subset. Since teams in the subset can only beat other teams in the subset, $x_i \le k-1$ for $i\le k$. For teams outside the subset, this condition guarantees they win at least $k$ games (as they beat everyone in the subset) and therefore, $x_i \ge k$ for $i > k$. Thus, if $y_k=\binom{k}{2}$, then $x_i \le k-1$ for $i\le k$ while $x_i \ge k$ for $i > k$. Therefore, if condition (2) is equality at $k$, then $x_k<x_{k+1}$. By negation, if $x_k = x_{k+1}$ (since $x_k$ cannot be greater than $x_{k+1}$) then condition (2) is a strict inequality at $k$. Consequently, if $x_{i} < x_{i+1} = x_{i+2} = ... = x_{i+m} < x_{i+m+1}$ then condition (2) is a strict inequality for $y_{i+1}$ through $y_{i+m-1}$ and the $U_{3,+}$ gate can be applied.

If we now consider a scoring sequence in which two teams differ by exactly two points then it is straightforward to show that the $U_{3,-}$ operation can be applied. Let $x_i< x_{i+1}=...=x_{i+m}<x_{i+m+1}$ with $x_{i}+2 = x_{i+m+1}$. Then, in this case, the order is preserved while $x_i$ is increased by one and $x_{i+m+1}$ is decreased by one. Further, the values $y_{i}$ through $y_{i+m}$ are all increased by one while the rest are unchanged. Thus, condition $(2)$ is trivially satisfied and the $U_{3,-}$ gate can be applied.

Since we know that $\{0,1,...,N-1\}$ corresponds to the state of one charge at every site, which belongs to the LKS, and the application of $U_{3,\pm}$ (the only operators needed to reach every state in the LKS) results in a valid scoring sequence, then every state in the LKS corresponds to a scoring sequence.

C) Consider an arbitrary scoring sequence. While there are two teams that have the same score, repeatedly apply the $U_{3,+}$ operation, which is always allowed by claim (B). The only case in which this process terminates is when the scoring sequence $\{0,1,...,N-1\}$ is reached. Since this scoring sequence corresponds to a state in the LKS, and since only 3-site gates were applied, it follows that the original scoring sequence must have also corresponded to a state in the LKS. Therefore, every valid scoring sequence corresponds to a state in the LKS.

With both (B) and (C), along with the fact that a scoring sequence cannot correspond to two different fracton states and vice versa, we prove that the number of valid scoring sequences for an $N$-team round robin tournament is equal to the number of state in the LKS for $N=L$.

\section{Labeling Krylov sectors by unique fully extended states}
\label{app:unique_configurations}

As part of the argument for the size of the LKS in Section \ref{sec:krylov}, we state that a Krylov sector can be uniquely labeled by its corresponding fully extended state, for which no $U_{3,+}$ operations are possible. Here, we will argue this point more strongly. 

Let us assume, for the sake of contradiction, that there exists a Krylov sector with arbitrary $N$, $L$, and $P$ that contains two fully extended states, $X$ and $Y$. We will denote the occupation numbers of the two fully extended states as $\{n_i^X\}$ and $\{n_i^Y\}$. 

Since $X$ and $Y$ belong to the same Krylov sector, we can create a sequence of $U_{3,\pm}$ operations that connect them. Let us imagine the sequence of operations the transforms $X$ into $Y$.
Suppose that the leftmost site in which $X$ and $Y$ differ is $x_a$ for some index $a$. 
If we assume, without loss of generality, that $n_a^Y > n_a^X$, then the extra particle at site $a$ must have been taken from sites $i > a$, since $X$ and $Y$ are identical at all smaller index. %(consequently bringing a particle from the left is not helpful since it would have to later be returned back from $i=a$). 
Therefore, the sequence of operations that transforms $X$ into $Y$ must include an operator $U_{3,+}$ applied at $i = a+1$, which brings the extra particle to site $i=a$. In order for this operation to be possible, there must have previously been two or more particles at site $i = a+1$. Since $n_{a+1}^{X}\le 1$ (by definition of a fully extended state), applying $U_{3,+}$ at $i = a+1$ requires again that a particle came from $i > a+1$. % (since, again, if the particle came from $i < a+1$, another $U_{3,+}$ would be required at $i = a+1$ to undo it). 
Repeating this logic, we see that constructing the state $Y$ apparently requires an operation $U_{3,+}$ to be applied at $i = L-2$, and thus $n_{L-2} \ge 2$. Since $X$ is fully extended, $n_{L-2}^{X} \le 1$, so applying $U_{3,+}$ at $i = L-2$ requires that a particle must have come from the right of it. However, this is impossible since no operation can be applied on the boundary ($i=L-1$). In short, the particle that arrives at site $i = a$ must be brought from the right, but by the definition of a fully extended state this extra particle can be taken neither from the bulk of the state nor from the rightmost boundary, and thus we arrive at a contradiction. So we must have $X = Y$. 

%This argument can be repeated for $n_a^Y < n_a^X$ to show that any two fully extended states cannot be reached from one another using only $U_{3,\pm}$. 
Therefore, there cannot be two fully extended states within a single Krylov sector, and consequently each Krylov sector can uniquely be labeled by its corresponding fully extended state.

\section{Generalizing the tournament analogy to arbitrary gate size}
\label{app:general_ell}

In Sec.~\ref{sec:krylov} we made an analogy between the size of the Krylov sector containing the uniform state and the number of scoring sequences in a round robin tournament. This analogy allowed us to prove that $n_c = 1$ for the case of gate size $\ell = 3$.
Here we consider the extension of this argument to generic gate size $\ell$. Specifically, we can make an analogy to the number of scoring sequences in an ($\ell-2$)-fold round robin tournament, in which each team plays every other team $\ell-2$ times. We will show that the number of scoring sequences in an $N$ team ($\ell-2$)-fold tournament is equal to the number of states in the Krylov sector that contains the state $X = \{1,0,..,0,1,0,...,0,1,0...,0,1,0,...,0,1\}$, where there are $N$ ones and $\ell-3$ zeros between each one.

The definition of a scoring sequence $\{x_1, x_2, ..., x_N\}$ for such a tournament is \cite{tournament}:
\begin{enumerate}[label={(\arabic*)}]
\item $0 \le x_1 \le x_2 \le ... \le x_{N} \le (\ell-2) (N-1)$
\item $y_k = \sum_{i=1}^{k} x_i \ge (\ell-2)\binom{k}{2}$.
\item $y_{N} = \sum_{i=1}^{N} x_i = (\ell-2)\binom{N}{2}$.
\end{enumerate}

We will now generalize the argument laid out in Appendix \ref{app:tournament_proof}. We begin by noticing that the scoring sequence $\{0, \ell-2, 2(\ell-2),...,(N-1)(\ell-2)\}$ satisfies these conditions and corresponds to the state we have called $X$. Therefore, we have a mapping between one state in the ($\ell-2$)-fold tournament and a state in the Krylov sector.

Next, it is clear that any $\ell$-site gate is equivalent to flipping the result of a particular game. Therefore, any sequence of $\ell$-site gates that transforms the state $X$ to a different state $Y$ in the Krylov sector can be represented as a series of game outcome flips that takes the fully extended scoring sequence to a different one. This mapping ensures that every state in the Krylov sector corresponds to a scoring sequence.

Finally, we will show that given a scoring sequence in which two teams scores differ by less than $\ell-2$, applying $U_{k\le\ell,+}$ results in a new valid scoring sequence. Begin by assuming we start with some scoring sequence $\{x_1,x_2,...x_N\}$ where $x_i + k = x_{i+m}$  with $0\le k < \ell-2 $ such that a $U_{k,+}$ gate can be applied to take $x_i \rightarrow x_i -1$ and $x_{i+m} \rightarrow x_{i+m} + 1$. Assume that there $\exists j$ with $i \le j < i+m$ such that $\sum_{i'=1}^j x_{i'}= y_j = (\ell-2)\binom{j}{2}$. This condition guarantees that teams one through $j$ only win games amongst themselves and lose all other games. Therefore,  since $i\le j$, $x_i \le (j-1)(\ell-2)$ and since $i+m > j$ $x_{i+m} \ge j(\ell-2)$. Consequently, we arrive at $x_{i+m}-x_{i} \ge \ell-2$ which contradicts the construction that $x_i + k = x_{i+m}$  with $0\le k < \ell-2 $. Therefore, by contradiction, we see that if two teams differ by $0\le k < \ell-2 $, then there is no $y_j = (\ell-2)\binom{j}{2}$ and thus applying $U_{k,+}$ will result in a valid scoring sequence satisfying (1)-(3).

This process will only terminate when all teams' scores differ by at least $\ell-2$. The resulting scoring sequence is unique, by the argument in Appendix \ref{app:unique_configurations}, and is precisely $X$. Therefore, all scoring sequences are reachable from $X$. This mapping ensures that every scoring sequence corresponds to a state in the Krylov sector.

Together, these arguments form a one-to-one mapping from states in the Krylov sector containing $X$ to the number of scoring sequences in an $N$-team $(\ell-2)$-fold tournament. 

Now we can turn to finding the number of scoring sequences in an $N$-team $(\ell-2)$-fold tournament. 

From the definition given by conditions (1)-(3) above, we follow the argument laid out in Ref.~\cite{moon_tournament}. First, we restrict ourselves to considering the case of $N=2M$. Then we can consider two sets of $M$ numbers that fully determine the original scoring sequence:

\begin{itemize}

\item $a_i = x_i$ for $1\le i \le M$
\item
$b_i = (\ell-2)(2M-1)-x_{2M+1-i}$ for $1\le i \le M$.

\end{itemize}

Here, the $a_i$ encode the first $M$ numbers in the scoring sequence and the $b_i$ encode the remaining $M$. With these new quantities, we can impose new constraints that imply conditions (1)-(3). %This weaker condition we will establish will provide a lower bound for the number of scoring sequences.
Counting the number of scoring sequences of this type provides a lower bound for the total number of scoring sequences.

Assume $a_i$ and $b_i$ meet the following constraints:
\begin{enumerate}[label={(\arabic*)}]
  \setcounter{enumi}{3}
\item $\sum_{i=1}^{M}a_i = \sum_{i=1}^{M}b_i$
\item $a_1 \le a_2 \le ... \le a_M = (\ell-2)(M-1)$
\item $a_k \ge (\ell-2)(k-1)$
\item $b_1 \le b_2 \le ... \le b_M = (\ell-2)(M-1)$
\item $b_k \ge (\ell-2)(k-1)$.
\end{enumerate}

From these constraints it is straightforward to show that (4) implies (3) while (5) and (6) imply (1). Additionally, constraint (6) implies (2) for  $k \le M$. Furthermore, conditions (7) and (3) imply (2) when $M+1 \le k \le 2M$.

Since conditions (4)-(8) imply (1)-(3), any valid sets of $a_i$ and $b_i$ will also correspond to a valid scoring sequence. Thus counting the number of valid sets of these new variables gives us a lower bound on the number of scoring sequences. 

Now we notice that if we denote $\sum_{i=1}^{M}a_i = T$, then the total number of sets fulfilling (4)-(8) is given by $Z$:
\begin{equation}
\label{eq:scoringseq}
    D_{\text{KS}}\left(n=\frac{1}{\ell-2};\ell,M \right) \ge Z = \sum_{T=1}^{(\ell-2)M^2} f(T;M,\ell)^2
\end{equation}
 where $f(T)$ is the number of sets of $a_i$ such that (4)-(6) are satisfied. The square comes from the fact that the $b_i$ are defined in exactly the same manner and have the same sum. Additionally, the sum, $T$ has to be between 1 and $(\ell-2)M^2$ (although many of those values will have $f(T;M,\ell)=0$).
 
 Next, we apply Jensen’s inequality to Eq.~\ref{eq:scoringseq} to obtain:
 \begin{equation}
\label{eq:scoringseq1}
    Z\ge \frac{1}{(\ell-2)M^2}\left(\sum_{T=1}^{(\ell-2)M^2} f(T;M,\ell)\right)^2.
\end{equation}

From this expression we notice that the summation is exactly the number of non-decreasing lattice paths from the origin to the point $(M-1, (\ell-2)(M-1))$ that stay at or above the line $y=(\ell-2)x$. This combinatorics problem, the ``weak ballot problem", is solved exactly by the $(\ell-2)$-Catalan numbers \cite{ballot_problem}:
 \begin{equation}
\label{eq:scoringseq2}
     \sum_{T=1}^{(\ell-2)M^2} f(T;M,\ell) = \frac{1}{(\ell-2)M+1}\binom{(\ell-1)M}{M}.
 \end{equation}
 
 Combining Eqs. (\ref{eq:scoringseq1}) and (\ref{eq:scoringseq2}), we obtain:
\begin{equation}
\label{eq:scoringseq3}
\begin{split}
     D_{\text{KS}} &\ge
     \frac{1}{(\ell-2)M^2}\left(\frac{1}{(\ell-2)M+1}\binom{(\ell-1)M}{M}\right)^2\\
     &\simeq \frac{(\ell-1)}{2 \pi(\ell-2)^3 \ell M^5}\left(\frac{(M(\ell-1))^{M(\ell-1)}}{M^M(M(\ell-2))^{M(\ell-2)}}\right)^2\\
     &\sim \frac{1}{M^5}\left(\frac{(\ell-1)^{(\ell-1)}}{(\ell-2)^{(\ell-2)}}\right)^{2M}.
\end{split}
\end{equation}
Remembering that $N=2M$ we arrive at:
\begin{equation}
 \label{eq:scoringseq4}
     D_{\text{KS}} \gtrsim \frac{1}{N^5}\left(\frac{(\ell-1)^{(\ell-1)}}{(\ell-2)^{(\ell-2)}}\right)^{N}.
 \end{equation}
 
Now that we have have a lower bound for the size of the Krylov sector, we can obtain an upper bound from the size of the symmetry sector. By construction, we have $L = (\ell-2)(N-1)+1$, which corresponds to $n=1/(\ell-2)$ in the limit of infinite system size. At this density, the size of the symmetry sector is given by Eq. \ref{eq:symmetry_sector_asym}:
 \begin{equation}
 \label{eq:upperbound}
\begin{split}
     D_{\text{sym}} &\sim
     \frac{1}{L^2}\left(\frac{(n+1)^{(n+1)}}{n^n}\right)^L\\
     &\sim\frac{1}{N^2}\left(\frac{\left(\frac{1}{\ell-2}+1\right)^{\left(\frac{1}{\ell-2}+1\right)}}{\frac{1}{\ell-2}^\frac{1}{\ell-2}}\right)^{(\ell-2)N}\\
     &\sim\frac{1}{N^2}\left(\frac{(\ell-1)^{(\ell-1)}}{(\ell-2)^{(\ell-2)}}\right)^{N}.
 \end{split}
 \end{equation}
 
With this expression, we now have an upper and lower bound on the size of the Krylov sector, both of which have the same exponential factor. Putting these together gives a relative size of the Krylov sector
 \begin{equation}
 \label{eq:upperbound_app}
     D_{\text{KS}} \sim\frac{1}{N^\alpha}\left(\frac{(\ell-1)^{(\ell-1)}}{(\ell-2)^{(\ell-2)}}\right)^{N}
 \end{equation}
with $2\le\alpha\le 5$. The lower bound on $D_{KS}$ can be tightened directly from Ref.~\cite{tournament} to $2\le\alpha\le 5/2$. From this expression, along with the conjecture that $D$ is exponentially small below $n_c$ and $1-D$ is exponentially small above $n_c$, we can extract the critical filling. It is clear that at $n=1/(\ell-2)$ the symmetry sector and a Krylov sector both have the same exponential scaling. Therefore, at this density, we have found a Krylov sector that makes up a power-law fraction of the symmetry sector and thus, $n_c=1/(\ell-2)$.

If we further conjecture that $\alpha = 5/2$ is constant in $\ell$ and that the Krylov sector we are considering is the largest one (using arguments similar to those in the main text), then we can obtain the scaling of $D$ for $n\le n_c$. With these assumptions, we arrive at:
 \begin{equation}
 \label{eq:D_general}
     D \sim\frac{n+1}{n^{3/2}\sqrt{L}}\left(\frac{\left(\frac{(\ell-1)^{(\ell-1)}}{(\ell-2)^{(\ell-2)}}n\right)^n}{(n+1)^{(n+1)}}\right)^{L}
 \end{equation}
for $n \le n_c = 1/(\ell-2)$.

\section{Algorithm for determining the size of largest Krylov sector}
\label{app:algo}

This section is largely based on Ref.~\onlinecite{ss_algo} and generalizes the result from $N=L$ to any $N$ and $L$ such that either $N\ge L$ or $(N+L) = 1\text{ }(\text{mod } 2)$. Reference~\onlinecite{ss_algo} presents a recursive algorithm for calculating the number of unique scoring sequence in an $N$-round robin tournament. It begins by defining a function $f_M(P,y)$ that counts the number scoring sequences $0 \le x_1 \le x_2 \le ... \le x_N = y$ with the following constraints:

\begin{enumerate}
\item \begin{equation}
    \sum_{i=1}^{N} x_i= P
\end{equation}
\item 
\begin{equation}
\label{condition}
\sum_{i=1}^k x_i \ge \binom{k}{2}.
\end{equation}

\end{enumerate}

Notice that $f_N(N(N-1)/2,N-1)$ is the number of scoring sequences for an $N$-team round robin tournament in which the maximum score is $N-1$. To find the total number of scoring sequences one can simply sum $f_N(N(N-1)/2,k)$ for $\lceil (N-1)/2\rceil \le k \le N-1$. This quantity can be calculated by now noticing the following recursive definition for $f_M(P,y)$: 
\begin{equation}
    f_1(P,y) = 
     \begin{cases} 
      1 & \text{if } P=y \geq 0 \\
      0 & \text{otherwise} 
   \end{cases}
\end{equation}
\begin{equation}
    f_M(P,y) = 
     \begin{cases} 
      \sum\limits_{k=0}^{P} f_{M-1}(P-y,k) &\text{if }  P-y \geq \binom{M-1}{2}\\
      0 & \text{otherwise.}
   \end{cases}
\end{equation}

This prescription allows for efficient calculation of the function $f_M(P,y)$ and therefore, the number of unique scoring sequences. We can then use our analogy between the number of scoring sequences and the size of the LKS in order to extend this algorithm to calculating the size of the LKS for any $N$ and $L$ such that either $N\ge L$ or $(N+L) = 1\text{ }(\text{mod } 2)$. In order to extend this formula, we notice that Eq.~\ref{condition} reaches equality for all $k$ exactly with the scoring sequence $\{0,1,2,...,N-1\}$. In the analogy to fractons, this scoring sequence corresponds to the uniform state. We notice that this is the fully extended state (see Appendix \ref{app:unique_configurations}). Further, since all states in the LKS can be reach from the uniform state through only the applications of $U_{k,-}$ operations, we are able to define a similar constraint based on this unique state (for all of the LKS for any $N$ and $L$ such that either $N\ge L$ or $(N+L) = 1\text{ }(\text{mod } 2)$). If we label the fully extended state in the LKS as $X = {X_1,X_2,...,X_N}$ (where $0\le X_i \le X_{i+1} \le N-1$), then the Eq.~\ref{condition} becomes 

\begin{equation}
\sum_{i=1}^k x_i \ge \sum_{i=1}^k X_i.
\end{equation}

Therefore, we can define a new equation $z(N,L,k) = \sum_{i=1}^{k}X_i$ based on the fully extended $X$ in the LKS with $N$ and $L$. Then, we have a new recursive function, $g_M(N,L,P,y)$, defined as:

\begin{equation}
    g_M(N,L,P,y) = 
     \begin{cases} 
      1 & \text{if } P = y \geq 0  \\
      0 & \text{otherwise} 
   \end{cases}
\end{equation}

\begin{equation}
\begin{split}
    g_M &(N,L,P,y) \\
    &= 
     \begin{cases} 
      \sum\limits_{k=0}^{P} g_{M-1}(P-y,k) &\text{if }  P-y \geq z(N,L,k)\\
      0 & \text{otherwise.} 
   \end{cases}
   \end{split}
\end{equation}

With this new function, we now have an efficient way in which to calculate number of states in the LKS:

\begin{equation}
    LKS(N,L) = 
     \sum_{k=0}^{N-1}g_N(N,L,N(L-1)/2,k).
\end{equation}

While this method will work generally for any $P$, we restrict ourselves to thinking about the case of $P=N(L-1)/2$ in which we have a conjectured structure of the fully extended state.

\section{Uniqueness of the fully extended state for periodic boundary conditions}
\label{app:fully_extended_PBC}

In this Appendix we consider the question of whether a fully extended state can exist, and whether it is unique, when the system has periodic boundary conditions. We focus on the case where the gate size $\ell = 3$.

A fully extended state is defined as a state for which no operations $U_{3,+}$ can be applied. For the case of closed boundary conditions, it is clear that a sequence of repeated applications of the operator $U_{3,+}$ must eventually terminate, since each application of $U_{3,+}$ increases the system's quadrupole moment, $Q = \sum_x n_x x^2$, by two. Therefore, since the fully extended state has a finite value of $Q$, there can be no infinite cycle of $U_{3,+}$ gates and thus a fully extended state must exist. However, when the system has periodic boundaries $Q$ can only be defined $(\text{mod } L^2)$, and thus the previous argument does not guarantee the existence of a fully extended state. In the remainder of this Appendix we demonstrate that any state with $N<L$ cannot be subjected to an infinite cycle of $U_{3,+}$ operators, and thus it must eventually reach a fully extended state consisting of only zeros and ones. We then show that this fully extended state is uniquely specified for a given starting state.

Consider an arbitrary starting state with some given values of $N$, $L$, and $P$ $(\text{mod } L)$. For the sake of contradiction, assume that there is an infinite set of $U_{3,+}$ gates that never reaches a fully extended state. Since there is a finite number of states, there must be some state $X$ that returns to itself after a finite set of gates have been applied. Let us denote the number of $U_{3,+}$ gates applied at site $x$ in during cycle as $a_x$. It is clear that since the charge is conserved during cycle, 
$$a_{x-1 (\text{mod} L)}-2 a_{x} + a_{x+1 (\text{mod} L)} = 0 \text{     for all }  x.$$

The only solution to this set of equations is $a_0 = a_1 = ... = a_{L-1} = a$. Therefore, any cycle consists of the same number of $U_{3,+}$ operations being applied at every site. Now, imagine marking a particle any time it is moved by one of these gates in the cycle. The first gate marks two particles. If the next gate is not applied to a site immediately adjacent to the first, then it will also mark two new particles, while if it is applied to a site adjacent to the first gate it will mark at least one new particle. In general, a gate will mark \emph{at least} two minus the number of gates previously applied adjacent to that site. Therefore, by the time a gate has been applied to every position, which is guaranteed whenever $a>0$, there are at least $L$ marked particles. Thus, any infinite cycle requires that $N\ge L$. Conversely, for $N<L$, the repeated application of $U_{3,+}$ gates must eventually terminate by producing a state for which no site $x$ has $n_x>1$, i.e., a fully extended state. Therefore, for the cases considered in Sec.~\ref{sec:PBC}, $N= L-1$ and $N = L-2$, a fully extended state is guaranteed to exist.

We can now prove that the fully extended state is unique, following a proof presented in Ref.~\cite{uniqueness_proof}. Consider an arbitrary initial state with $N<L$ and assume for the sake of contradiction that there are two different sequences of $U_{3,+}$ gates that reach different fully extended states.
%, for which the maximum occupation in a given state is one. 
We can denote the sequences of $U_{3,+}$ operations by $X =\{x_1, x_2, ..., x_n\}$ and $Y = \{y_1, y_2, ..., y_m\}$, where $x_i$ ($y_i$) denote the position of the $i$-th gate in the first (second) set of gates. Since the two sequences reach different final states starting from the same initial state, there must be some first instance, $k$, where $x_k \neq y_k$. Since a gate can be applied at $x_k$ to the state reached after the first $k-1$ gates, there must be at least two particles at $x_k$. Since the final state will not have two particles at any site, there must be some future gate such that $k'$ is the next instance in $Y$ where $y_{k'} = x_k$ to ensure that this site eventually reaches less than two. So despite not applying the gate to $x_k$ at the $k$-th step, there must be some future step $k'$ that does.

Since this gate is applied later in the $Y$ sequence, we can modify the order of this sequence by moving $y_{k'}$ to the $k$-th position. This move changes the $Y$ sequence to $Y' = {y_1,...y_{k-1}, y_{k'}, y_{k}, y_{k+1},...,y_{k'-1},y_{k'+1},...,y_{m}}$. This new sequence is still valid since we know that $y_{k'} = x_{k}$ is a valid gate to apply at step $k$ since $Y$ is the same as $X$ up to this point. Further, since there are no $y_{i} = y_{k'}$ for $k \le i < k'$, and all other positions have a charge greater than equal to what its value would have been without the application of $y_{k'}$, the remaining gates are allowed. Therefore, $Y'$ contains all the same gates as $Y$, and so reaches the same final state, but is now identical to $X$ up to at least the first $k+1$ gates. By repeating this procedure of swapping the order of gate application in $Y$, we will eventually reach a point when the new sequence is equal to $X$. At this point it is clear that $X$ and $Y$ must be the same up to the order of the gates applied. Therefore, the fully extended states that they reach must be identical. 

These two proofs together establish that a system with periodic boundary conditions has a unique fully extended state any time $N < L$.
In fact, the arguments presented here can be equally applied to the case of closed boundary conditions, which recovers the result presented in Appendix \ref{app:unique_configurations}.

\section{Algorithm for selecting a random state}
\label{app:random_state}

Here we present an algorithm for selecting a random state from the symmetry sector with a particular charge $N$ and dipole moment $P$.  If we consider periodic boundary conditions, then generating a random state for a system with $\text{GCD}(N,L)=1$ can be done as follows: 
\begin{enumerate}
    \item Draw a random variable, $N_0$, which corresponds to the number of particles at the first site of the system. $N_0$ is drawn from the probability distribution $p(N_0) = \binom{N-N_0+L-2}{L-2}/\binom{N+L-1}{L-1}$ which comes from the number of states for the remaining $N-N_0$ particles on the $L-1$ sites divided by all possible states. Place $N_0$ particles at the first site.
    \item Consider the remaining sites and particles of the system, i.e., update $N\rightarrow N-N_0$ and $L\rightarrow L-1$.
    \item Repeat steps 1 and 2 until $N$ and/or $L$ reaches 0.
    \item The resulting state has some dipole moment $P$. If we shift all of the particles one space to the right, then we will increase the dipole moment by, $N$ which is guaranteed to change the dipole moment since $P$ is defined modulo $L$. Continue shifting the state until the desired $P$ is reached.
\end{enumerate}

\section{Derivation of the correlation length exponent}
\label{app:nu}

\begin{figure}[tb!]
\begin{center}
\includegraphics[width=0.9\columnwidth]{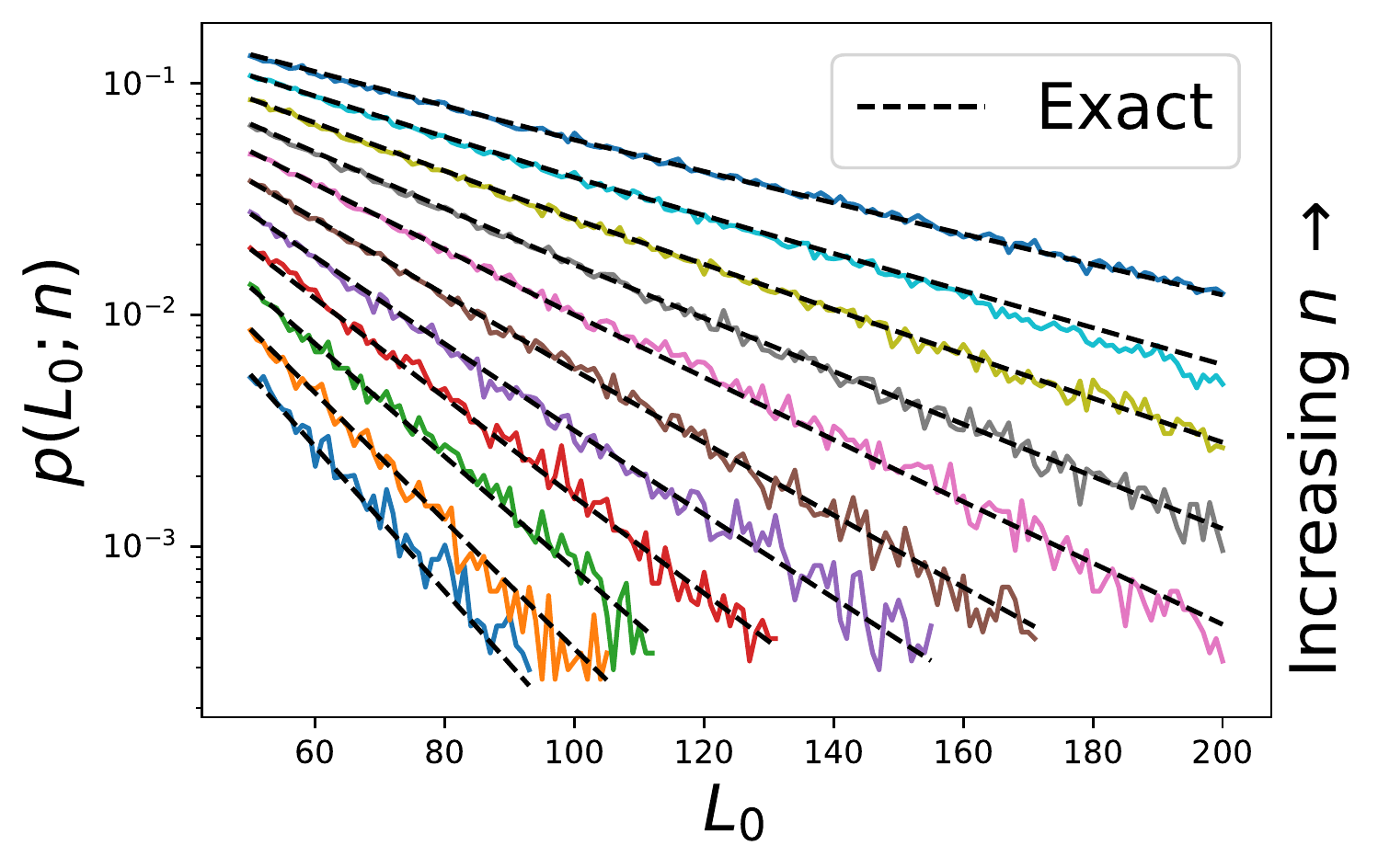}
\end{center}
    \caption{Numerically-estimated probability that a segment of length $L_0$ segment has particle density $n\ge n_c$. We generate 375 random states with $L=10000$ with values of the average density $n$ ranging from 0.6 (bottom curve) to 0.8 (top curve). For each curve, we sample 100 random segments of length $L_0$ and calculate the probabilty that the segment is locally thermalized (its density exceeds $n_c=1$). Our numerical results closely match the theoretical result of Eq.~\ref{eq:prob}, and they display exponential decay with $L_0$ in the range $1\ll L_0 \ll L$. %Ultimately the form of this probability allows us to obtain an expression for $\xi$.
    }
    \label{fig:numerical_nu}
\end{figure}
In the main text we give a heuristic argument for the correlation length exponent $\nu = 2$. Here we make that argument more rigorous. To do so, we  calculate the probability that a given segment of a state is locally thermalized, i.e., it has a filling that exceeds the critical value. If there is a well defined correlation length $\xi$, this probability should decay exponentially with the size of the segment, and the exponential decay constant defines $\xi$. As before, we will assume that $\text{GCD}(N,L)= 1$. This assumption assures that to create a random state (detailed in Appendix \ref{app:random_state}) one can ignore the dipole contstraint and then circularly shifting the origin of the coordinate axis. Therefore, the probability $p(n, L_0)$ that a segment of length $L_0$ contains at least $N_0=n_c L_0$ particles is given by
\begin{equation}
\label{eq:prob}
 p(n,L_0) = \sum_{N_0 = n_c L_0}^{N} \frac{\binom{N_0 + L_0 -1}{L_0-1}\binom{N-N_0+L-L_0-1}{L-L_0-1}}{\binom{N+L-1}{L-1}}.
\end{equation}

In Fig.~\ref{fig:numerical_nu} we compare this expression with results obtained from numerical simulations, which show strong agreement. Equation (\ref{eq:prob}) can be rearranged to obtain:
\begin{equation}
    \label{eq:prob2}
    \begin{split}
    &p(n,L_0)\\
    &= \frac{N!(L-1)!}{(L_0-1)!(L-L_0-1)!(N+L-1)!}\\
    &\times \sum_{N_0=n_c L_0}^{N}\frac{(N_0+L_0-1)!((n+1)L-N_0-L_0-1)!}{N_0!(N-N_0)!}\\
    &= \frac{N!(L-1)!((n_c+1)L_0-1)!((n+1)L-(n_c+1)L_0-1)!}{(L_0-1)!(L-L_0-1)!(N+L-1)!(n_c L_0)!(nL-n_c L_0)!}\\
    &\times \sum_{N_0'=0}^{N-n_c L_0}\prod_{i=1}^{N_0'}\frac{((n_c+1)L_0-1+i)(nL-n_cL_0-i)}{(n_cL_0+i)((n+1)L-(n_c+1)L_0-1-i)}.
    \end{split}
\end{equation}
Roughly, the expression inside the summation on the last line contributes most when $N_0'$, and thus $i$, is small with respect to $nL-n_c L$. Therefore, the product simplifies to $(n(n_c+1)/(n_c(n+1)))^{N_0'}$ in the limit of $1\ll L_0 \ll L$ and fixed $n$. Therefore,
\begin{equation}
    \label{eq:prob3}
    \begin{split}
    &p(n,L_0) \simeq\\
    &\frac{N!(L-1)!((n_c+1)L_0-1)!((n+1)L-(n_c+1)L_0-1)!}{(L_0-1)!(L-L_0-1)!((n+1)L-1)!(n_c L_0)!(nL-n_c L_0)!}\\
    &\times \sum_{N_0'=0}^{N-n_c L_0}\left(\frac{n(n_c+1)}{n_c(n+1)}\right)^{N_0'}\\
    &\simeq \frac{N!L!((n_c+1)L_0)!((n+1)L-(n_c+1)L_0)!}{(L_0)!(L-L_0)!((n+1)L)!(n_c L_0)!(nL-n_c L_0)!}\\
    &\times\frac{L_0(L-L_0)((n+1)L)}{L(n_c+1)L_0((n+1)L-(n_c+1)L_0)}\frac{n_c(n+1)}{n_c-n}\\
    &\simeq \frac{N!L!((n_c+1)L_0)!((n+1)L-(n_c+1)L_0)!}{(L_0)!(L-L_0)!((n+1)L)!(n_c L_0)!(nL-n_c L_0)!}\\
    &\times\frac{n_c(n+1)}{(n_c+1)(n_c-n)}.
    \end{split}
\end{equation}

Through several applications of Stirling's approximation and subsequently taking the limit where $L$ and $L_0$ go to infinity we arrive at
\begin{equation}
    \label{eq:prob4}
    \begin{split}
    &p(n,L_0)\\ 
    &\simeq \frac{n_c(n+1)}{(n_c+1)(n_c-n)}\sqrt{\frac{nL(n_c+1)((n+1)L-(n_c+1)L_0)}{2\pi L_0 (L-L_0)(n+1)n_c(nL-n_c L_0)}}\\
    &\times \text{exp}\left(-\frac{nL(n_c+1)((n+1)L-(n_c+1)L_0)}{L_0 (L-L_0)(n+1)n_c(nL-n_c L_0)}\right) \\
    &\times \left(\frac{n^n ((n+1)L - (n_c+1)L_0)^{(n+1)}}{(n+1)^{(n+1)}(L-L_0)(nL-n_cL_0)^n}\right)^L\\
    &\times \left(\frac{(n_c+1)^{(n_c+1)}(L-L_0)(nL-n_cL_0)^{n_c}}{n_c^{n_c}((n+1)L-(n_c+1)L_0)^{n_c+1}}\right)^{L_0}\\
    &\simeq \frac{n+1}{n_c-n}\sqrt{\frac{n_c}{2\pi (n_c+1)L_0}}\left(\frac{(n_c+1)^{(n_c+1)}n^{n_c}}{n_c^{n_c}(n+1)^{n_c+1}}\right)^{L_0}\\
    &= \frac{n+1}{n_c-n}\sqrt{\frac{n_c}{2\pi (n_c+1)L_0}} e^{\text{log}\left(\frac{(n_c+1)^{(n_c+1)}n^{n_c}}{n_c^{n_c}(n+1)^{n_c+1}}\right)L_0}\\
    &\equiv \frac{n+1}{n_c-n}\sqrt{\frac{n_c}{2\pi (n_c+1)L_0}} e^{-L_0/\xi}.
    \end{split}
\end{equation}

This last expression defines the correlation length by equating the exponential factor in the expression for $p(n, L_0)$ with $\text{exp}[-L_0/\xi]$. Therefore, we arrive at
\begin{equation}
    \label{eq:xi}
    \begin{split}
    \xi(n)  &\simeq \frac{-1}{\text{log}\left(\frac{(n_c+1)^{(n_c+1)}n^{n_c}}{n_c^{n_c}(n+1)^{n_c+1}}\right)}\\
    &\simeq \frac{2n_c(n_c+1)}{(n-n_c)^2}.
    \end{split}
\end{equation}
This last equality corresponds to the limit $n_c - n \ll n_c$, and recovers the expected result $\nu=2$.

\def\bibsection{\section*{\refname}} 

\bibliography{fracton.bib}

\clearpage

\end{document}